\PassOptionsToPackage{bookmarks=false}{hyperref}
\documentclass[11pt]{article}

\usepackage[preprint]{acl}

\usepackage{times}
\usepackage{latexsym}

\usepackage[T1]{fontenc}

\usepackage[utf8]{inputenc}

\usepackage{microtype}

\usepackage{inconsolata}

\usepackage{graphicx}
\usepackage{amsmath}
\usepackage{amssymb}
\usepackage{booktabs}
\usepackage{enumitem}
\usepackage{algorithm}
\usepackage{algpseudocode}
\usepackage{float}
\usepackage{caption}
\usepackage[most]{tcolorbox}
\usepackage{listings}
\graphicspath{{../}}
\AtBeginDocument{\hfuzz=10000pt\vfuzz=10000pt\hbadness=10000\vbadness=10000}

\lstdefinestyle{promptstyle}{
  basicstyle=\ttfamily\scriptsize,
  breaklines=true,
  columns=fullflexible,
  keepspaces=true,
  showstringspaces=false,
  frame=none,
  xleftmargin=0pt,
  xrightmargin=0pt
}

\newtcolorbox{promptbox}[1]{
  enhanced,
  breakable,
  colback=gray!3,
  colframe=gray!55,
  boxrule=0.35pt,
  arc=1.5pt,
  left=4pt,
  right=4pt,
  top=3pt,
  bottom=3pt,
  title={#1},
  fonttitle=\bfseries\footnotesize,
  coltitle=black
}

\usepackage{multirow}
\usepackage{booktabs}

\title{All-Mem: Agentic Lifelong Memory via Dynamic Topology Evolution}

\author{
  \textbf{Can Lv\textsuperscript{1,2}},
  \textbf{Heng Chang\textsuperscript{3}},
  \textbf{Shengyu Tao\textsuperscript{4}},
  \textbf{Mingju Chen\textsuperscript{1,2}},
\\
  \textbf{Zhaoxin Fan\textsuperscript{1,2}},
  \textbf{Ziwei Zhang\textsuperscript{2}},
  \textbf{Yuchen Guo\textsuperscript{3}},
  \textbf{Shiji Zhou\textsuperscript{1,2}}
\\
\\
  \textsuperscript{1}Beijing Advanced Innovation Center for Future Blockchain and Privacy Computing\\
  \textsuperscript{2}Beihang University
  \textsuperscript{3}Tsinghua University
  \textsuperscript{4}Chalmers University of Technology\\
\\
  \small{
    \textbf{Project Lead:} Heng Chang,
    \textbf{Corresponding to:} Shiji Zhou
    \href{mailto:zhoushiji25@buaa.edu.cn}{\texttt{<zhoushiji25@buaa.edu.cn>}}
  }
}

\begin{document}
\maketitle
\begin{abstract}
Lifelong interactive agents are expected to assist users over months or years, which requires continually writing long term memories while retrieving the right evidence for each new query under fixed context and latency budgets. Existing memory systems often degrade as histories grow, yielding redundant, outdated, or noisy retrieved contexts.
We present \textbf{All-Mem}, an online/offline lifelong memory framework that maintains a topology structured memory bank via explicit, non destructive consolidation, avoiding the irreversible information loss typical of summarization based compression.
In online operation, it anchors retrieval on a bounded visible surface to keep coarse search cost bounded.
Periodically offline, an LLM diagnoser proposes confidence scored topology edits executed with gating using three operators: Split, Merge, and Update, while preserving immutable evidence for traceability.
At query time, typed links enable hop bounded, budgeted expansion from active anchors to archived evidence when needed.
Experiments on \textbf{LoCoMo} and \textbf{LongMemEval-s} show improved retrieval and QA over representative baselines.
The code is available at \url{https://github.com/LvCan926/All-Mem}.
\end{abstract}

\section{Introduction}
\label{sec:intro}

Lifelong autonomous agents must continually accumulate experiences, decisions, and user-specific facts~\cite{towardslifelonglearning, zheng2025lifelongagentbenchevaluatingllmagents, maharana2024evaluatinglongtermconversationalmemory}. As this history grows without bound, the central challenge becomes \emph{selective evidence retrieval} under fixed context and latency budgets~\cite{kang2025memoryosaiagent, MemoryBank, chhikara2025mem0buildingproductionreadyai}.
In practice, naive accumulation progressively contaminates the retrievable pool: entangled units, redundant variants, and superseded versions
increasingly occupy a fixed budget, yielding diffuse or even misleading contexts for downstream reasoning~\cite{generativeagents, lostinthemiddle,wu2025longmemevalbenchmarkingchatassistants}.

\begin{figure}[t]
  \centering
  \includegraphics[width=\linewidth]{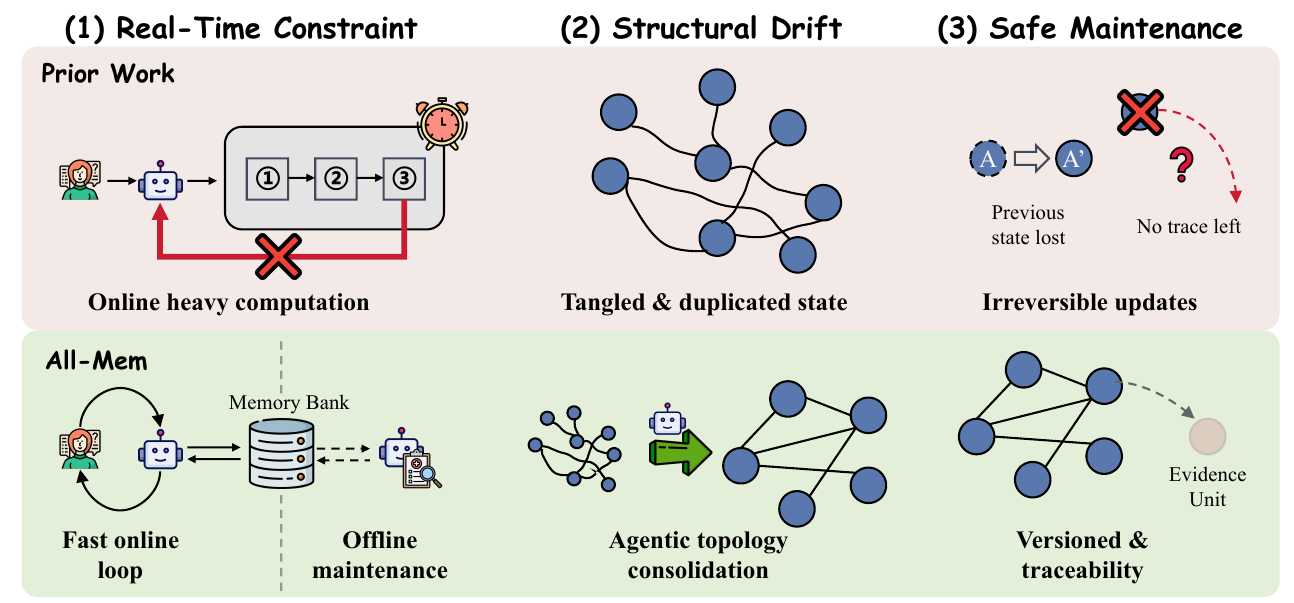}
  \caption{Comparison of prior works and All-Mem along real-time constraints, structural drift, and safe maintenance (traceability).}
  \label{fig:intro}
  \vspace{-6pt}
\end{figure}

Existing agent memory systems span runtime context orchestration and compression/structure-assisted retrieval beyond flat similarity search~\cite{edge2025localglobalgraphrag, sarthi2024raptorrecursiveabstractiveprocessing,guo2025lightragsimplefastretrievalaugmented}.
Recent work also introduces LLM-assisted maintenance to rewrite, merge, or suppress stored items~\cite{xu2025amemagenticmemoryllm, chhikara2025mem0buildingproductionreadyai}.
However, long-horizon deployments still lack a unified maintenance-and-recovery interface: updates are often performed via \emph{in-place} replacement or ad-hoc archiving, without explicit lineage and \emph{budgeted} recovery to immutable evidence, so redundancy, entanglement, and superseded states increasingly crowd a fixed window~\cite{MemoryBank, chhikara2025mem0buildingproductionreadyai,rasmussen2025zeptemporalknowledgegraph}.

Despite this progress, lifelong memory pipelines repeatedly face three root-cause challenges.
\textbf{(1) Real-Time Constraint.}
Meaningful re-organization requires cross-item context and nontrivial computation, yet the online interaction loop must remain low latency~\cite{zhou2025mem1learningsynergizememory,chen2026iterresearchrethinkinglonghorizonagents}. This tension makes reliable maintenance difficult to perform within the same online budget~\cite{fang2025lightmemlightweightefficientmemoryaugmented,modarressi2024retllmgeneralreadwritememory}.
\textbf{(2) Structural Drift.}
Over long horizons, memory organization inevitably accumulates structural debt. Items become conflated, duplicated, or left in unresolved state transitions~\cite{zheng2025lifelongagentbenchevaluatingllmagents,wang2023voyageropenendedembodiedagent}. When the system relies on fixed organization rules, fixed segmentation, or append-only growth, these artifacts increasingly crowd a fixed retrieval budget, causing evidence to become diffuse, outdated, or internally inconsistent~\cite{packer2024memgptllmsoperatingsystems,sun2025scalinglonghorizonllmagent}.
\textbf{(3) Safe Maintenance.}
Correcting drift requires structural edits that may have global side effects~\cite{hu2026memoryageaiagents}. Updates therefore must be verifiable and reversible with preserved provenance, and they must retain a reliable path back to immutable evidence, otherwise maintenance is either applied too weakly to matter or applied irreversibly with hidden errors~\cite{hu2023chatdbaugmentingllmsdatabases,pan2024kwaiagentsgeneralizedinformationseekingagent}.

To address these challenges, we propose \textbf{All-Mem}, an \textbf{agentic lifelong memory framework} inspired by Complementary Learning Systems (CLS)~\cite{o2014complementary}, which couples fast online acquisition with offline re-organization.
All-Mem has three components:
\textbf{(1) Online/Offline Decoupling} keeps the interaction loop low-latency by limiting online writing to lightweight ingestion with sparse, revisable links and by restricting coarse search to a curated \emph{visible surface}.
\textbf{(2) Agentic Topology Consolidation} runs offline to counter structural drift using \emph{non-destructive} edits: an LLM diagnoser proposes targets, confidence-gating filters unreliable actions, and an LLM planner regenerates descriptors before deterministic execution of Split, Merge, and Update with versioned traceability to immutable evidence.
\textbf{(3) Topology-Aware Retrieval} preserves recoverability without exposing the full archive by anchoring on the visible surface and performing hop-limited, budgeted expansion along typed links when finer-grained evidence is needed.

Our contributions are:
\begin{itemize}
    \item We propose \textbf{All-Mem}, an \emph{online/offline} agentic lifelong memory framework, and first introduce \emph{Agentic Topology Consolidation} as a \emph{non-destructive} consolidation paradigm for lifelong LLM agents.

    \item We develop a non-destructive topology editing mechanism (Split, Merge, Update) and a topology-guided retrieval mechanism with budgeted expansion that preserves recoverability to immutable evidence.

    \item Across long-horizon memory benchmarks, All-Mem consistently improves retrieval quality and downstream question answering over strong accumulative baselines as interaction histories grow.
\end{itemize}

\section{Related Work}
\label{sec:related}

\paragraph{Long-Horizon Memory for LLM Agents.}
Long-running LLM agents extend finite context windows with external read--write memory, where the central challenge is to decide what to store, compress, update, and retrieve under fixed latency and context budgets.
Prior systems address this challenge through segmentation~\cite{MemoryBank, comedychen2024compressimpressunleashingpotential}, summarization~\cite{yu2025memagentreshapinglongcontextllm}, prompt or memory compression~\cite{jiang2023llmlinguacompressingpromptsaccelerated, chevalier2023adaptinglanguagemodelscompress,tan2025prospectretrospectreflectivememory}, forgetting~\cite{liu2023thinkinmemoryrecallingpostthinkingenable, li2025memosoperatingmemoryaugmentedgeneration}, memory updating~\cite{wang2023voyageropenendedembodiedagent, chhikara2025mem0buildingproductionreadyai}, and runtime retrieval orchestration~\cite{fang2025lightmemlightweightefficientmemoryaugmented, packer2024memgptllmsoperatingsystems}.
For example, MemGPT manages information across active and external context through virtual context management~\cite{packer2024memgptllmsoperatingsystems}, while Mem0 extracts, updates, and retrieves salient memories, with graph-enhanced variants for relational modeling~\cite{chhikara2025mem0buildingproductionreadyai}.
These methods improve long-context interaction by optimizing memory selection, compression, and retrieval-time context construction.
However, as histories grow, memory banks also accumulate \emph{structural drift}: redundant variants, conflated units, and superseded states that continue to compete for the same retrieval budget~\cite{maharana2024evaluatinglongtermconversationalmemory, wu2025longmemevalbenchmarkingchatassistants}.
All-Mem addresses this complementary problem by explicitly maintaining which memories remain searchable and how archived evidence remains recoverable.

\paragraph{Agentic Memory Maintenance.}
Recent work introduces adaptive maintenance mechanisms that revise, filter, or reorganize memories as new interactions arrive.
Examples include memory updating for embodied or conversational agents~\cite{wang2023voyageropenendedembodiedagent, chhikara2025mem0buildingproductionreadyai}, selective forgetting or operating-memory policies~\cite{liu2023thinkinmemoryrecallingpostthinkingenable, li2025memosoperatingmemoryaugmentedgeneration}, and offline or sleep-time consolidation to reduce online latency~\cite{fang2025lightmemlightweightefficientmemoryaugmented}.
These approaches keep long-term memory useful under growing histories, but their maintenance operations are usually formulated as extraction, rewriting, summarization, filtering, or task-specific update policies.
They do not explicitly expose memory maintenance as a topology-level interface for repairing conflation, redundancy, and supersession while preserving lineage to immutable evidence.

A-Mem is the closest work to ours: it builds an agentic, Zettelkasten-style memory network by enriching new memories with contextual descriptions, keywords, and tags, dynamically linking related memories, and evolving existing memory representations~\cite{xu2025amemagenticmemoryllm}.
All-Mem is complementary but targets a different stage and failure mode.
Whereas A-Mem emphasizes memory construction, representation evolution, and link generation, All-Mem focuses on post-hoc topology repair of an accumulated memory bank.
It separates three maintenance decisions that are often entangled in prior systems: which units should remain visible, which units should be archived, and how archived evidence remains recoverable.
Concretely, All-Mem defines explicit \textsc{Split}, \textsc{Merge}, and \textsc{Update} operators to handle conflated, redundant, and superseded memory units.
These edits are \emph{non-destructive}: raw evidence is never overwritten, source or superseded units are archived through a visibility indicator, and typed links preserve recoverability to the original evidence.
Thus, memory evolution in All-Mem not only enriches representations, but also safely removes noisy or outdated units from the online searchable surface while keeping them traceable.

\paragraph{Structured Retrieval and Memory.}
Another line of work organizes retrieval corpora as explicit structures, such as graphs, trees, or hierarchies, to support evidence aggregation beyond flat nearest-neighbor search~\cite{gutiérrez2025ragmemorynonparametriccontinual, edge2025localglobalgraphrag, sarthi2024raptorrecursiveabstractiveprocessing, hu2024hiagenthierarchicalworkingmemory, guo2025lightragsimplefastretrievalaugmented}.
RAPTOR recursively clusters and summarizes text chunks into a tree for multi-level retrieval~\cite{sarthi2024raptorrecursiveabstractiveprocessing}; HippoRAG-style systems use graph structures and multi-hop traversal to improve associative retrieval and sense-making~\cite{gutiérrez2025ragmemorynonparametriccontinual}; and LightRAG improves the efficiency of graph-based indexing and retrieval~\cite{guo2025lightragsimplefastretrievalaugmented}.
These methods show that explicit structure can improve retrieval when evidence is distributed across passages.

However, most structured-retrieval methods treat structure primarily as a retrieval index or abstraction layer over a corpus, rather than as a persistent memory state that must be actively maintained under continual interaction.
They therefore do not directly address how to repair conflated, duplicated, or superseded memories without losing provenance.
All-Mem couples structured retrieval with explicit topology evolution: offline ATC periodically applies confidence-gated non-destructive edits, while query-time retrieval anchors on a curated visible surface and performs hop-bounded expansion over typed links.
This makes the retrieval structure itself a maintained object, preserving bounded online cost and recoverability to archived immutable evidence.

\begin{figure*}[t]
  \centering
  \includegraphics[width=\textwidth,height=0.33\textheight,keepaspectratio]{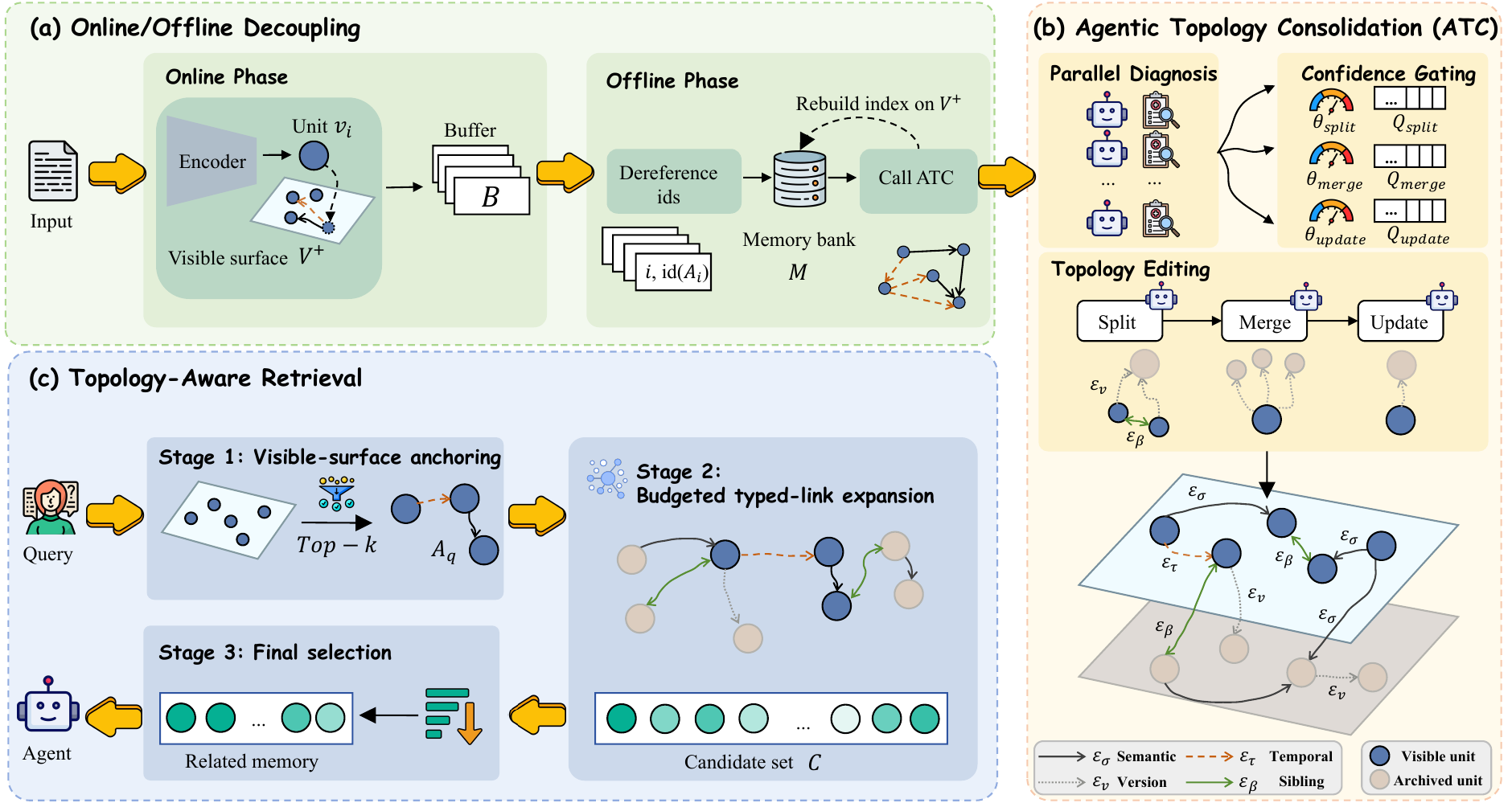}
  \caption{All-Mem framework. (a) Online/offline decoupling with id-level buffering. (b) ATC performs parallel diagnosis with confidence-gating and non-destructive topology edits (Split$\rightarrow$Merge$\rightarrow$Update) while preserving versioned traceability. (c) Topology-aware retrieval anchors on the visible surface, performs budgeted typed-link expansion, and selects final memories for the agent.}
  \label{fig:framework}
  \vspace{-8pt}
\end{figure*}

\section{Methodology}
\subsection{Problem Formulation}
\label{subsec:formulation}

We model long-term memory at dialogue step $N$ as a topology-structured memory bank
$\mathcal{M}_N=(\mathcal{V}_N,\mathcal{E}_N)$, where $\mathcal{V}_N$ stores memory units and
$\mathcal{E}_N$ stores typed directed links for controllable maintenance and retrieval.
Each unit $v_i\in\mathcal{V}_N$ is
\begin{equation}
    v_i=\langle c_i, s_i, \mathcal{K}_i, \mathbf{z}_i, t_i, a_i\rangle,
\end{equation}

where $c_i$ is immutable raw evidence, $s_i$ is a regenerable summary, $\mathcal{K}_i$ is a regenerable keyword set,
and $\mathbf{z}_i$ is the embedding of the string concatenation of $s_i$ and $\mathcal{K}_i$, $t_i$ is a timestamp, and $a_i\in\{0,1\}$ is a revisable visibility indicator.

\paragraph{Visible surface.}
We restrict vector search to the \emph{visible} subset
\begin{equation}
\mathcal{V}^{+}_{N}=\{\, v_i\in\mathcal{V}_{N}\mid a_i=1\,\},
\end{equation}
thereby decoupling the coarse retrieval domain from the unbounded history size.
Deactivation ($a_i\leftarrow 0$) is non-destructive archiving: raw evidence $c_i$ is never deleted, and archived units remain
recoverable via typed links.

\paragraph{Typed-link topology and hop-bounded recoverability.}
We organize links into four families,
$\mathcal{E}_{N}=\mathcal{E}_{\tau}\cup \mathcal{E}_{\sigma}\cup \mathcal{E}_{\nu}\cup \mathcal{E}_{\beta}$,
corresponding to temporal continuity, semantic association (with an out-degree cap), non-destructive versioning (new$\rightarrow$old), and sibling coherence.
Let $\mathrm{dist}_{\mathcal{E}}(u,v)$ denote the directed shortest-path hop distance in the subgraph induced by edge set $\mathcal{E}$.
We maintain the invariant that every archived unit is reachable from some visible unit; define $H_N$ such that
\begin{equation}
\forall v \in \mathcal{V}_{N}\setminus \mathcal{V}^{+}_{N},\ \exists u \in \mathcal{V}^{+}_{N}
\ \text{s.t.}\ \mathrm{dist}_{\mathcal{E}_{N}}(u,v)\le H_{N}.
\end{equation}

\subsection{Online/Offline Decoupling}
\label{subsec:decoupling}

\paragraph{Online phase}
At interaction step $N$, the agent performs \emph{minimal online writing} and records id-level pointers for deferred \emph{offline consolidation}.
Online writing consists of three actions:
(i) Unit Writing (creating a new unit and generating its descriptor fields),
(ii) Surface Linking (sparsely attaching it to the \emph{visible} surface), and
(iii) Buffering (recording \emph{id-level} metadata for offline processing).
As a result, the per-turn cost depends on nearest-neighbor search over the visible subset
$\mathcal{V}^{+}_{N}$ rather than over the full bank, plus a single LLM call for descriptor generation.

Given an observation $c$ at step $N$, we create a new unit $v_i$ (where $i$ is a unit identifier and is not tied to $N$)
with immutable evidence $c_i\!\leftarrow\!c$, regenerate a summary $s_i$ and keywords $\mathcal{K}_i$, and embed their concatenation to obtain
the index vector $\mathbf{z}_i$.

We attach $v_i$ using sparse, provisional links that target only the visible surface.
Let $v_{\mathrm{prev}}$ denote the session-local predecessor (if available), and retrieve a small anchor set
\begin{equation}
\mathcal{A}_i=\operatorname{Top-}k\big(\mathbf{z}_i,\mathcal{V}^{+}_{N}\big),
\end{equation}
where Top-$k$ is computed under cosine similarity.
When $v_{\mathrm{prev}}$ exists, we insert a temporal link $(v_i, v_{\mathrm{prev}})\in\mathcal{E}_{\tau}$.
We additionally add a few degree-capped semantic attachments $(v_i,u)\in\mathcal{E}_{\sigma}$ for $u\in\mathcal{A}_i$.
These online links are provisional bootstraps and may be revised or superseded by offline consolidation; expensive topology edits
(e.g., Split/Merge/Update) are deferred.

Finally, we push the new unit id $i$ and neighbor ids $\mathrm{ids}(\mathcal{A}_i)$ into the consolidation buffer $\mathcal{B}$ for offline processing.
The buffer $\mathcal{B}$ therefore contains pointers rather than duplicated evidence or descriptor fields, keeping online overhead minimal.

\paragraph{Offline Phase}
Outside the interaction loop, the system periodically consumes buffered id records in $\mathcal{B}$ and updates the memory bank state
$\mathcal{M}=(\mathcal{V},\mathcal{E})$ to improve the searchability of the visible surface while preserving non-destructive traceability to
immutable evidence.

Since $\mathcal{B}$ stores only identifiers, the offline worker first dereferences each buffered record to instantiate a local context
$x_b$ (the unit descriptors and a Top-$k$ anchor neighborhood). It then invokes \emph{Agentic Topology Consolidation}
(ATC; Sec.~\ref{subsec:agentic_topology_consolidation}) on the set of contexts $\mathcal{X}=\{x_b\}_{b\in\mathcal{B}}$ to produce bank updates
(unit Split/Merge/Update, link rewiring, and visibility-indicator updates $a_i$ within affected units), which are applied
back to $\mathcal{M}$.

After applying updates, we reindex the memory bank by regenerating embeddings for modified descriptors and rebuilding the search index
over the visible subset $\mathcal{V}^{+}_{N}$.
This keeps per-turn retrieval dependent on the maintained visible surface, while preserving typed-link traceability to archived evidence.

\begin{algorithm}[t]
\caption{Agentic Topology Consolidation}
\label{alg:offline}
\begin{algorithmic}[1]
\Require Memory bank $\mathcal{M}=(\mathcal{V},\mathcal{E})$, contexts $\mathcal{X}=\{x_b\}$
\Ensure Updated memory bank $\mathcal{M}=(\mathcal{V},\mathcal{E})$
\ForAll{$op\in\{Split,Merge,Update\}$}
  \State $\mathcal{Q}_{op}\gets \emptyset$
\EndFor

\State \textbf{Parallel diagnosis + routing}
\ForAll{$x\in\mathcal{X}$}
  \ForAll{$(op,T^{*},p)\in \mathrm{LLM}\!\left(P_{\mathrm{diag}}(x)\right)$}
  \If{$p\ge \theta_{op}$}
    \State $\mathcal{Q}_{op}\gets \textsc{Norm}(\mathcal{Q}_{op}\cup\{T^{*}\};\mathcal{M})$
  \EndIf
\EndFor
\EndFor

\State \textbf{Serial editing}
\ForAll{$op\in[Split,Merge,Update]$}
  \ForAll{$T^{*}\in\mathcal{Q}_{op}$}
  \If{\textsc{Applicable}$(T^{*};\mathcal{M})$}
    \State $R \gets \mathrm{LLM}\!\left(P_{\mathrm{plan}}(op,T^{*};\mathcal{M})\right)$
    \State $(\mathcal{V},\mathcal{E})\gets \textsc{ApplyPlan}(R;\mathcal{V},\mathcal{E})$
  \EndIf
\EndFor
\EndFor
\end{algorithmic}
\end{algorithm}

\subsection{Agentic Topology Consolidation}
\label{subsec:agentic_topology_consolidation}

ATC consolidates buffered ids to curate the visible surface while preserving non-destructive traceability to immutable evidence
(Algorithm~\ref{alg:offline}; Fig.~\ref{fig:framework}b). It consists of:
(i) \emph{Parallel diagnosis} over Top-$k$ anchor neighborhoods,
(ii) \emph{confidence-gating} into operator-specific queues, and
(iii) \emph{Topology editing} via deterministic, degree-constrained archiving and rewiring.

\paragraph{Parallel diagnosis.}
For each context $x\in\mathcal{X}$, an LLM diagnoser proposes candidate topology edits as
\begin{equation}
\mathcal{P}(x)=\{(op,T^{*},p)\},
\end{equation}
where $op\in\{\textsc{Split},\textsc{Merge},\textsc{Update}\}$ is an operator, $T^{*}$ is an operator-specific target, and
$p\in[0,1]$ is the diagnoser confidence.
Specifically, $T^{*}=v$ for \textsc{Split}, $T^{*}=S$ for \textsc{Merge}, and $T^{*}=(u,v)$ for \textsc{Update}, where $v$ is a candidate conflated unit, $S$ is a redundant visible set, and $(u,v)$ is an ordered current--superseded pair.
Proposals are obtained by querying the LLM with a fixed diagnosis template:
\begin{equation}
\mathcal{P}(x)\leftarrow \mathrm{LLM}\!\left(P_{\mathrm{diag}}(x;\mathcal{M})\right).
\end{equation}
Diagnosis over different $x\in\mathcal{X}$ is embarrassingly parallel.

\paragraph{Confidence-gating.}
We accept a proposal $(op,T^{*},p)$ if $p\ge \theta_{op}$, where $\theta_{op}$ is a fixed operator-specific threshold tuned on a held-out
development set. Accepted targets are normalized (e.g., deduplicated, id-sorted, and filtered) and routed into operator-specific queues
$\mathcal{Q}_{\textsc{Split}},\mathcal{Q}_{\textsc{Merge}},\mathcal{Q}_{\textsc{Update}}$ for subsequent editing. This confidence-gating limits
unreliable edits while keeping diagnosis and routing lightweight and parallelizable.

Topology editing. Given queued targets, we execute accepted edits serially
under a fixed operator order SPLIT$\rightarrow$MERGE$\rightarrow$UPDATE,
which prioritizes disentangling, then deduplication, and finally
version-lineage refinement. At execution time, we run an applicability check
to skip stale or conflicting targets. For \textsc{Split}, the LLM planner
proposes grounded evidence segments, and the executor derives sibling
descriptors using the descriptor compression prompt. For \textsc{Merge} and
\textsc{Update}, the LLM planner regenerates the canonical or refreshed
descriptor. All archiving and link rewiring are executed deterministically
under degree constraints to preserve typed-link traceability and prevent
densification.

\paragraph{Non-destructive operators.}
ATC uses three operators to repair the dominant forms of structural drift; full rewiring rules are in Appendix~\ref{app:operators}.
\textsc{Split} decomposes a conflated unit $v$ into visible siblings $S_v=\{v'_1,\dots,v'_m\}$, archives $v$, and connects siblings and source evidence through $\mathcal{E}_{\beta}$ and $\mathcal{E}_{\nu}$.
\textsc{Merge} consolidates a redundant visible set $S\subseteq\mathcal{V}^{+}$ into a new representative $\tilde v$, archives the sources, and rewires a degree-capped subset of external links to $\tilde v$.
\textsc{Update} refreshes the descriptor of a current unit $u$, archives its superseded predecessor $v$, and adds a version link $(u,v)\in\mathcal{E}_{\nu}$.
In all cases, raw evidence remains immutable, archiving only toggles visibility ($a_i\leftarrow0$), and modified descriptors are re-embedded during reindexing.

\begin{table*}[t]
\centering
\caption{\textbf{Performance on \textbf{LoCoMo} and \textbf{LongMemEval-s}.}
Answer quality: 4o-J/F1/BLEU-1/ROUGE-L; retrieval: R@5/N@5.
\textit{Caveat:} LoCoMo uses turn-level ID matching, while LongMemEval-s uses session-level matching, so retrieval scores are not directly comparable across datasets.}
\label{tab:unified_results}
\small
\setlength{\tabcolsep}{3pt}

\begin{tabular}{l|cccc|cc @{\hspace{6pt}} cccc|cc}
\toprule
& \multicolumn{6}{c}{\textit{\textbf{LoCoMo}}} & \multicolumn{6}{c}{\textit{\textbf{LongMemEval-s}}} \\
\cmidrule(r){2-7}\cmidrule(l){8-13}
\textbf{Method} &
\textbf{4o-J} & \textbf{F1} & \textbf{B-1} & \textbf{R-L} & \textbf{R@5} & \textbf{N@5} &
\textbf{4o-J} & \textbf{F1} & \textbf{B-1} & \textbf{R-L} & \textbf{R@5} & \textbf{N@5} \\
\midrule
Full History & 41.39 & 29.57 & 26.76 & 29.03 & - & - & 47.00 & 15.30 & 17.18 & 25.62 & - & - \\
Naive RAG & 37.94 & 32.56 & 29.05 & 33.94 & 28.08 & 26.16 & 45.80 & 24.92 & 16.76 & 25.11 & 71.83 & 69.57 \\
MemGPT~\cite{packer2024memgptllmsoperatingsystems} & 33.19 & 25.34 & 20.51 & 25.74 & 24.47 & 21.04 & 42.80 & 20.25 & 12.84 & 20.70 & 70.24 & 67.43 \\
A-Mem~\cite{xu2025amemagenticmemoryllm} & 40.74 & 34.59 & 30.14 & 34.56 & 31.94 & 28.75 & 50.40 & 30.82 & 22.77 & 30.91 & 81.36 & 78.24 \\
HippoRAG 2~\cite{gutiérrez2025ragmemorynonparametriccontinual} & 42.37 & 34.79 & 29.67 & 33.56 & 34.95 & 33.19 & 53.20 & 32.90 & 24.35 & 32.80 & 84.13 & 81.76 \\
Mem0~\cite{chhikara2025mem0buildingproductionreadyai} & 48.91 & 43.08 & 40.63 & 45.21 & 38.74 & 32.13 & 55.80 & 36.10 & 28.05 & 36.06 & 90.17 & 87.14 \\
LightMem~\cite{fang2025lightmemlightweightefficientmemoryaugmented} & 44.39 & 41.94 & 37.13 & 42.23 & 37.04 & 35.78 & 54.20 & 34.33 & 26.28 & 34.12 & 87.42 & 85.48 \\
\textbf{All-Mem (Ours)} & \textbf{54.63} & \textbf{52.18} & \textbf{46.31} & \textbf{52.01} & \textbf{46.63} & \textbf{41.02} &
\textbf{60.20} & \textbf{45.19} & \textbf{36.42} & \textbf{45.94} & \textbf{94.68} & \textbf{93.27} \\
\bottomrule
\end{tabular}
\vspace{-6pt}
\end{table*}

\subsection{Topology-Aware Retrieval}
\label{subsec:retrieval}

We perform topology-aware retrieval with a coarse-to-fine pipeline.
Instead of relying solely on flat vector similarity over the full bank, we first retrieve anchors from the \emph{visible} surface,
then expand along typed links under explicit hop and candidate budgets, and finally re-rank the bounded candidate set for precision.

\paragraph{Stage 1: Visible-surface anchoring.}
Given a query $q$, we compute its embedding $\mathbf{z}_q$ and retrieve $k$ anchors from the visible surface:
\begin{equation}
\mathcal{A}_q=\operatorname{Top-}k\big(\mathbf{z}_q,\mathcal{V}^{+}_{N}\big).
\end{equation}

\paragraph{Stage 2: Budgeted typed-link expansion.}
We expand from $\mathcal{A}_q$ along typed links with hop limit $H_q$ to form a bounded candidate set:
\begin{equation}
\mathcal{C}=\textsc{Expand}(\mathcal{A}_q;H_q,L),\quad |\mathcal{C}|\le L,
\end{equation}
using a fixed priority over link types (favoring versioning/sibling coherence).

\paragraph{Stage 3: Final selection.}
We rank candidates in $\mathcal{C}$ by cosine similarity between $\mathbf{z}_q$ and the cached unit embedding $\mathbf{z}_i$,
and materialize a memory context by attaching evidence $c_i$ for the selected units. The resulting context is provided to the agent
as retrieval-augmented input to answer $q$.

Overall, the retrieval procedure is explicitly budgeted by $(k,H_q,L)$: $k$ visible anchors, hop limit $H_q$, and at most $L$ expanded candidates,
so all reranking and context materialization costs scale with $L$.

\section{Experiments}
\label{sec:experiments}

\subsection{Experimental Setup}
\label{subsec:setup}

\paragraph{Benchmarks and metrics.}
We evaluate long-horizon conversational memory on \textbf{LoCoMo}~\cite{maharana2024evaluatinglongtermconversationalmemory} and \textbf{LongMemEval-s}~\cite{wu2025longmemevalbenchmarkingchatassistants}.
We report retrieval metrics (R@5, N@5) and answer-quality metrics (4o-J\cite{NEURIPS2023_91f18a12}, F1, BLEU-1, ROUGE-L).

\paragraph{Baselines.}
We compare against \emph{Full History}, a dense-retrieval baseline (\emph{Naive RAG}), and memory-centric methods
(\emph{MemGPT}~\cite{packer2024memgptllmsoperatingsystems}, \emph{A-Mem}~\cite{xu2025amemagenticmemoryllm}, \emph{HippoRAG2}~\cite{gutiérrez2025ragmemorynonparametriccontinual},
\emph{Mem0}~\cite{chhikara2025mem0buildingproductionreadyai}, \emph{LightMem}~\cite{fang2025lightmemlightweightefficientmemoryaugmented}).
All methods follow the same incremental protocol and share the same generator; we additionally match retrieval and context budgets
via an identical input-token cap to the generator.

\paragraph{Implementation Details.}
We use \emph{GPT-4o-mini} (temperature $=0$) as the generator,
\emph{All-MiniLM-L6-v2}~\cite{minilm} for embeddings. All methods share the same generator and are compared under matched retrieval and context budgets.

\begin{figure}[t]
    \centering
    \includegraphics[width=\columnwidth]{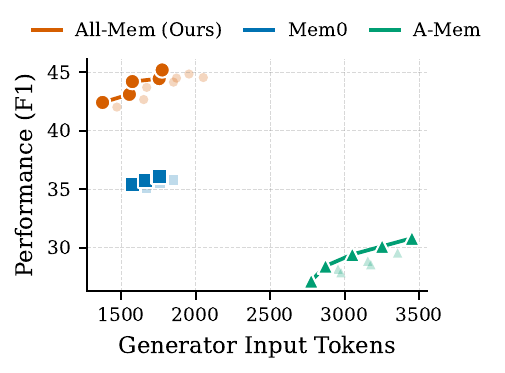}
    \caption{\textbf{Accuracy--tokens sweep on \textbf{LongMemEval-s}.}
Each point is one retrieval-budget setting. All-Mem and A-Mem vary $(K,k,L)$; Mem0 varies $(K,k)$.}
    \label{fig:pareto_tokens_longmemeval}
    \vspace{-6pt}
\end{figure}

\begin{figure*}[t]
    \centering
    \includegraphics[width=\textwidth]{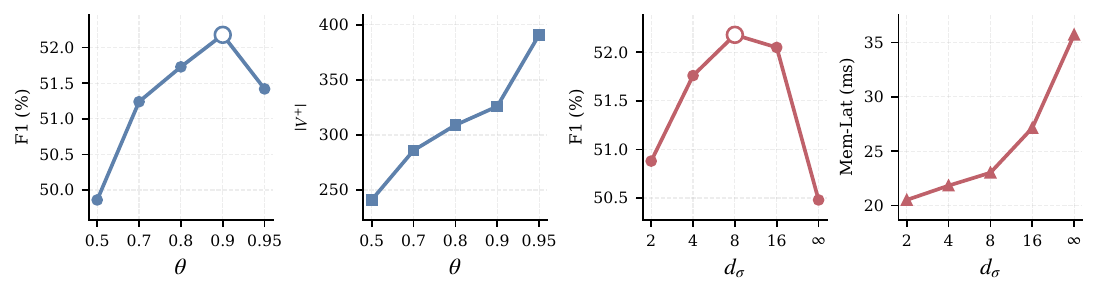}
    \caption{\textbf{Sensitivity analysis on \textbf{LoCoMo}.}
    Left: varying the ATC threshold $\theta$ affects F1 and visible-surface size $|V^{+}|$.
    Right: varying the semantic out-degree cap $d_{\sigma}$ affects F1 and memory-module latency.
    Defaults $\theta{=}0.9$ and $d_{\sigma}{=}8$ give the best trade-off.}
    \label{fig:sensitivity}
    \vspace{-6pt}
\end{figure*}

\subsection{Main Results}
\label{subsec:main_results}

Table~\ref{tab:unified_results} summarizes answer quality and retrieval quality on LoCoMo and LongMemEval-s.
\textbf{All-Mem} achieves the best performance across both benchmarks, consistently improving downstream QA and evidence retrieval over prior memory-centric baselines.

On LoCoMo, All-Mem obtains 54.63 4o-J and 52.18 F1, outperforming the strongest prior baseline Mem0 by 5.72 and 9.10 points, respectively.
It also improves turn-level retrieval, reaching 46.63 R@5 and 41.02 N@5; compared with the best prior retrieval scores, this corresponds to gains of 7.89 R@5 over Mem0 and 5.24 N@5 over LightMem.
These results suggest that topology consolidation improves not only answer generation but also fine-grained evidence localization.

On LongMemEval-s, All-Mem achieves 60.20 4o-J and 45.19 F1, improving over Mem0 by 4.40 and 9.09 points.
Retrieval also improves from 90.17/87.14 R@5/N@5 with Mem0 to 94.68/93.27 with All-Mem.
Since LongMemEval-s uses session-level retrieval matching and retrieval scores can saturate, we interpret these retrieval gains together with the larger downstream QA improvements.

\subsection{Budgeted Retrieval: Accuracy--Tokens Pareto Sweep}
\label{subsec:pareto_tokens}

We evaluate the accuracy--cost trade-off on LongMemEval-s by sweeping the retrieval budgets while keeping all other components fixed.
Specifically, $k$ controls the number of visible-surface anchors, $L$ controls the size of typed-link expansion, and $K$ controls the number of final memories materialized into the generator context.
We use realized generator input length as the deployment cost proxy, since it directly reflects the query-time context injected into the LLM.

As shown in Fig.~\ref{fig:pareto_tokens_longmemeval}, All-Mem forms a stronger F1--token frontier than the memory baselines.
Under comparable token budgets with Mem0 (approximately 1.4k--2.1k tokens), All-Mem consistently achieves higher F1, showing that visible-surface anchoring and budgeted expansion improve evidence quality rather than merely reducing prompt length.
A-Mem operates in a higher-token regime (roughly 2.8k--3.5k tokens), but remains below All-Mem in F1, suggesting weaker cost-effectiveness under the same accounting.
Increasing $L$ generally improves F1 with diminishing returns, and the best All-Mem setting in our sweep is $(K{=}16,k{=}10,L{=}40)$, reaching 45.19 F1.

\subsection{Sensitivity Analysis}
\label{subsec:sensitivity}

We test the robustness of All-Mem to two key hyperparameters on LoCoMo: the shared ATC confidence threshold $\theta$, which is used uniformly for Split, Merge, and Update proposals, and the semantic-link out-degree cap $d_{\sigma}$, which controls the density of semantic expansion during retrieval.

Fig.~\ref{fig:sensitivity} shows that All-Mem remains stable near the default configuration, but both parameters expose clear quality--cost trade-offs.
For $\theta$, lower thresholds accept more topology edits across all three operators and reduce the visible-surface size, but excessive consolidation can remove useful active anchors and slightly hurt F1.
Conversely, overly high thresholds make ATC conservative: more units remain visible, including redundant or superseded memories, which also reduces retrieval quality.
The best F1 is obtained at $\theta{=}0.9$, with nearby settings remaining competitive.

For $d_{\sigma}$, small caps under-connect the memory topology, limiting typed-link expansion and reducing evidence coverage.
Increasing $d_{\sigma}$ initially improves F1 by allowing more useful semantic neighborhoods, but overly large caps introduce dense expansion, increasing memory-module latency without further accuracy gains.
The default $d_{\sigma}{=}8$ therefore gives the best quality--cost trade-off.

\subsection{Cost Analysis}
\label{subsec:cost_analysis}

We break down the cost of All-Mem into online writing, query-time retrieval, and offline ATC consolidation.
All measurements are collected on LongMemEval-s under the same setting.
As shown in Table~\ref{tab:cost_overall}, the online path remains lightweight: online writing costs 2.38s and 539 tokens per turn, while query-time retrieval costs 27.54ms and injects 918 tokens per query.
Offline ATC is more expensive per event, but it is paid periodically and amortizes to 0.21s and 1237 tokens per turn.

Table~\ref{tab:operator_cost} decomposes offline ATC cost.
Diagnosis dominates token usage because the diagnoser must inspect many local contexts before routing candidate edits.
Among executed edits, Merge accounts for the largest event-level cost because it is the most frequent operation, while Split has higher procedural complexity because it requires evidence grouping and descriptor regeneration.
Overall, the profile supports the intended online/offline separation: heavy topology maintenance is moved out of the interaction loop, while query-time retrieval remains bounded by the visible surface and expansion budget.

\begin{table}[t]
\centering
\footnotesize
\setlength{\tabcolsep}{5pt}
\caption{
\textbf{Overall cost profile of All-Mem on LongMemEval-s.}
Online/offline tokens count LLM input+output tokens for memory construction or maintenance; retrieval tokens denote mean generator input tokens per query.
}
\begin{tabular}{lcc}
\toprule
\textbf{Component} 
& \textbf{Latency} 
& \textbf{Tokens} \\
\midrule
Online writing        & 2.38s / turn    & 539 / turn \\
Offline ATC event     & 12.79s / event  & 74.1k / event \\
Amortized offline     & 0.21s / turn    & 1237 / turn \\
Query-time retrieval  & 27.54ms / query & 918 / query \\
\bottomrule
\end{tabular}
\label{tab:cost_overall}
\vspace{-8pt}
\end{table}

\begin{table}[t]
\centering
\footnotesize
\setlength{\tabcolsep}{4pt}
\caption{
\textbf{Offline ATC cost decomposition on LongMemEval-s.}
Counts report total diagnoses, executed edits, or offline events over the full run; event-level costs are averaged over offline ATC events.
}
\begin{tabular}{lccc}
\toprule
\textbf{Component}
& \textbf{Total Count}
& \textbf{Time / Event}
& \textbf{Tok. / Event} \\
\midrule
Diagnosis    & 621k   & 1.57s  & 48.2k \\
Split        & 12.6k  & 1.77s  & 4.0k \\
Merge        & 87.9k  & 7.77s  & 18.2k \\
Update       & 25.1k  & 1.68s  & 3.8k \\
\midrule
Edits total  & 125.6k & 11.22s & 25.9k \\
\textbf{ATC total} & 8.37k & \textbf{12.79s} & \textbf{74.1k} \\
\bottomrule
\end{tabular}
\label{tab:operator_cost}
\vspace{-8pt}
\end{table}

\subsection{Ablation Study}
\label{subsec:ablation}

We conduct ablations on LoCoMo to isolate three core design choices: visible-surface gating, offline topology consolidation, and budgeted recovery.
Table~\ref{tab:ablation_selected} reports representative variants; the full ablation suite is in Appendix~\ref{app:ablation_full}.

\paragraph{Visible-surface gating.}
No-Visibility forces all units to remain searchable by setting $a_i{=}1$.
This reduces F1 from 52.18 to 47.63 and R@5 from 46.63 to 41.84, while increasing memory-module latency from 23.02ms to 34.58ms.
Thus, exposing the full archive harms both selectivity and efficiency.

\paragraph{Offline consolidation.}
Removing a key topology-edit operator, e.g., w/o Merge, lowers F1 by 3.16 points and R@5 by 2.67 points.
This shows that consolidation is not only a storage-cleaning step: removing redundant variants improves the quality of the visible surface under fixed retrieval budgets.

\paragraph{Budgeted recovery.}
Anchors-only disables typed-link expansion after Stage~1 anchoring.
Although it is faster, F1 drops from 52.18 to 49.27 and R@5 drops from 46.63 to 43.19, indicating that anchors alone miss useful supporting evidence.
No-recovery-links further removes traversal over recovery links, reducing F1 to 50.06.
These results confirm that hop-bounded expansion and recovery topology are necessary for balancing selective retrieval with archived-evidence recoverability.

\begin{table}[t]
\centering
\footnotesize
\setlength{\tabcolsep}{4pt}
\caption{Selected ablations on LoCoMo. Mem-Lat is mean memory-module latency (Stage~1--3; excluding generator inference). Full ablations are in Appendix~\ref{app:ablation_full} (Table~\ref{tab:ablation_full}).}
\begin{tabular}{l|cccc}
\toprule
\textbf{Variant} & \textbf{4o-J} $\uparrow$ & \textbf{F1} $\uparrow$ & \textbf{R@5} $\uparrow$ & \textbf{Mem-Lat} $\downarrow$ \\
\midrule
\textbf{All-Mem (Full)} & \textbf{54.63} & \textbf{52.18} & \textbf{46.63} & 23.02 \\
No-Visibility & 50.87 & 47.63 & 41.84 & 34.58 \\
w/o Merge & 52.08 & 49.02 & 43.96 & 27.40 \\
Anchors-only & 51.76 & 49.27 & 43.19 & \textbf{14.27} \\
No-recovery-links & 52.74 & 50.06 & 44.07 & 22.67 \\
\bottomrule
\end{tabular}
\label{tab:ablation_selected}
\vspace{-8pt}
\end{table}

\section{Conclusion}
We introduced All-Mem, a lifelong memory framework that frames long-horizon agent memory as a \emph{budgeted maintenance-and-recovery} problem. It maintains a topology-structured bank with a curated visible surface, applies confidence-gated non-destructive consolidation, and recovers archived evidence through hop-bounded typed-link expansion. This online/offline design keeps retrieval bounded while allowing the memory bank to be repaired as histories grow. Across LoCoMo and LongMemEval-s, All-Mem improves retrieval and downstream QA; further analyses show the gains come from better surface maintenance and budgeted recovery rather than larger context injection. Future work includes adaptive operator scheduling, richer link semantics, and privacy-aware retention/deletion.

\section{Limitations}
\label{sec:limitations}

All-Mem is designed as a safe and auditable memory-maintenance framework, but several limitations remain.

ATC depends on LLM-based diagnosis and planning to determine when topology consolidation is needed and how it should be applied. This design gives the framework flexibility across heterogeneous memory states, but the quality of topology edits is still bounded by the reliability of the underlying model. Inaccurate diagnoses may lead to overly fine-grained splits, overly aggressive merges, or imperfect temporal-supersession judgments. All-Mem mitigates these risks through confidence gating, executor-side validation, fail-closed execution, and non-destructive archiving: raw evidence is never overwritten or deleted, and archived units remain recoverable through typed links. Nevertheless, erroneous edits may still affect which units are exposed on the visible retrieval surface. Future work can further improve edit reliability through calibrated confidence estimation, verifier models, or ensemble-based diagnosis.

All-Mem also shifts part of the memory-management cost from the online interaction loop to offline consolidation. This keeps query-time retrieval bounded and low-latency, but assumes that background maintenance is available. Deployments with very frequent updates, limited background compute, or strict energy constraints may require different consolidation schedules. In practice, the consolidation interval, buffer size, and operator thresholds provide natural control knobs for adapting All-Mem to different resource budgets.

The current implementation uses a compact operator set, consisting of \textsc{Split}, \textsc{Merge}, and \textsc{Update}. These operators cover common forms of structural drift, including conflation, redundancy, and supersession, but they do not exhaust all memory-maintenance needs. Real-world agents may additionally require explicit deletion, user-controlled retention, access control, domain-specific schemas, or task-conditioned memory policies. Similarly, the typed-link topology in All-Mem is specified by design rather than learned end-to-end, which may limit adaptability in some domains. Extending the operator set and learning richer link semantics are promising directions.

The empirical evaluation focuses on two long-horizon conversational memory benchmarks. These benchmarks evaluate core retrieval and question-answering abilities under growing interaction histories, but they do not cover the full range of lifelong-agent deployments, such as multimodal memory, collaborative multi-user memory, safety-critical environments, or long-term personalization under evolving user preferences. The results should therefore be interpreted as evidence for the studied long-horizon conversational setting, rather than as a complete validation of all lifelong-memory scenarios.

Preserving archived evidence improves provenance, traceability, and recoverability, but it also raises privacy and data-minimization considerations in practical deployments. A production memory system should incorporate explicit retention, deletion, and access-control mechanisms so that non-destructive maintenance remains aligned with user expectations and applicable policies.

\bibliography{custom}

\appendix

\clearpage

\begin{center}
    \Large\bfseries Appendix
\end{center}

\section{Operator Semantics}
\label{app:operators}

This appendix provides the executable semantics of the three topology operators used by Agentic Topology Consolidation (ATC). 
The LLM only produces a structured repair plan $R$; all graph edits, validation, archiving, and rewiring are performed by a deterministic executor.
Invalid or stale plans are rejected and executed as no-ops.

\paragraph{Common state.}
A memory unit $v$ contains immutable raw evidence $c_v$, regenerable descriptors $(s_v,\mathcal{K}_v)$, an embedding $\mathbf{z}_v$, a timestamp $t_v$, and a visibility indicator $a_v\in\{0,1\}$.
The visible surface is
\begin{equation}
\mathcal{V}^{+}=\{v\in\mathcal{V}:a_v=1\}.
\end{equation}
Archiving a unit only changes its visibility, i.e., $a_v\leftarrow 0$; the raw evidence $c_v$ is never deleted or overwritten.
The edge set contains the typed links defined in the main text:
$\mathcal{E}=\mathcal{E}_{\tau}\cup\mathcal{E}_{\sigma}\cup\mathcal{E}_{\nu}\cup\mathcal{E}_{\beta}$,
corresponding to temporal, semantic, version, and sibling-coherence links.

\subsection{\textsc{Split}}
\label{app:operators:split}

Input. \textsc{Split} targets a visible unit $v \in V^+$ whose evidence contains
multiple separable facts or topics. The repair plan $R_{\textsc{Split}}$
contains a list of grounded evidence segments. If the list contains fewer than
two non-empty segments, the edit is treated as a no-op.

Execution. If the plan is valid and contains $m \ge 2$ grounded segments, the executor creates a visible sibling set
\[
S_v = \{v'_1,\ldots,v'_m\}, \quad a_{v'_j}=1.
\]
Each sibling $v'_j$ receives one segment of the original evidence as immutable
evidence. Its descriptor $(s_{v'_j}, K_{v'_j})$ is generated by the descriptor
compression prompt.

\subsection{\textsc{Merge}}
\label{app:operators:merge}

\paragraph{Input.}
\textsc{Merge} targets a redundant visible set $S\subseteq\mathcal{V}^{+}$ with $|S|\ge2$.
The repair plan $R_{\textsc{Merge}}$ provides a canonical descriptor for a new representative node.

\paragraph{Execution.}
If the target set and plan pass validation, the executor creates a new visible representative $\tilde{v}$:
\begin{equation}
a_{\tilde{v}}=1.
\end{equation}
The representative stores references to the source units in $S$ and uses the plan-generated descriptor $(s_{\tilde{v}},\mathcal{K}_{\tilde{v}})$.
All source units are archived:
\begin{equation}
a_v\leftarrow 0,\qquad \forall v\in S.
\end{equation}
The executor adds version links from the representative to each archived source:
\begin{equation}
(\tilde{v},v)\in\mathcal{E}_{\nu},\qquad \forall v\in S.
\end{equation}
A degree-capped subset of semantic links incident to the archived sources is redirected to $\tilde{v}$.
This keeps the visible surface compact while preserving access to the original evidence through version links.

\subsection{\textsc{Update}}
\label{app:operators:update}

\paragraph{Input.}
\textsc{Update} targets an ordered pair $(u,v)$, where $u\in\mathcal{V}^{+}$ is the current unit and $v$ is a superseded unit.
The repair plan $R_{\textsc{Update}}$ refreshes the descriptor of $u$ using its evidence and local context.

\paragraph{Execution.}
If the plan is valid, the executor replaces the descriptor of the current unit:
\begin{equation}
(s_u,\mathcal{K}_u)\leftarrow R_{\textsc{Update}}.
\end{equation}
The current unit is marked for re-embedding.
The superseded unit is archived if it is still visible:
\begin{equation}
a_v\leftarrow 0.
\end{equation}
The executor adds a version link from the current unit to the superseded unit:
\begin{equation}
(u,v)\in\mathcal{E}_{\nu}.
\end{equation}
If semantic links previously pointed to $v$, the executor may redirect a degree-capped subset of them to $u$.

\subsection{Deterministic Rewiring and Validation}
\label{app:operators:rewiring}

\paragraph{Executor-side validation.}
Before execution, the executor checks basic structural constraints: referenced ids must exist, required units must be visible, operator arity must be valid, and descriptor fields must be non-empty.
For \textsc{Split}, each proposed segment must be grounded in the source evidence.
For \textsc{Merge}, the target set must contain at least two visible units and pass the configured redundancy check.
For \textsc{Update}, the current unit must be visible.
Any failed check causes the target to be skipped.

\paragraph{Degree-capped rewiring.}
Semantic and sibling links are maintained under fixed out-degree caps.
When a unit $x$ is archived, the executor defines a visible representative set $\mathrm{Rep}(x)$:
\begin{equation}
\mathrm{Rep}(x)=
\begin{cases}
S_x, & \text{if }x\text{ is split},\\
\{\tilde{v}\}, & \text{if }x\text{ is merged},\\
\{u\}, & \text{if }x\text{ is superseded by }u.
\end{cases}
\end{equation}
For an incoming semantic edge $(p,x)\in\mathcal{E}_{\sigma}$ from a visible node $p$, the executor redirects it to the best available representative:
\begin{equation}
r^{*}
=
\arg\max_{r\in \mathrm{Rep}(x)}
\mathrm{sim}(\mathbf{z}_p,\mathbf{z}_r),
\end{equation}
with deterministic tie-breaking by timestamp and node id.
After insertion, outgoing neighborhoods are pruned to satisfy the configured degree caps.

\section{ATC Execution Protocol}
\label{app:atc_protocol}

This appendix specifies how Agentic Topology Consolidation (ATC) turns LLM diagnosis outputs into deterministic topology edits.
The protocol follows the same high-level procedure as Algorithm~\ref{alg:offline}: diagnosis is parallel, while editing is serial and fail-closed.
The goal is to use the LLM for proposal and descriptor generation, but keep all state changes deterministic.

\subsection{Diagnosis Proposals}
\label{app:atc:diagnosis}

For each local context $x\in\mathcal{X}$, the diagnoser outputs a set of candidate edit proposals:
\begin{equation}
\mathcal{P}(x)=\{(op,T^{*},p)\},
\end{equation}
where $op\in\{\textsc{Split},\textsc{Merge},\textsc{Update}\}$ is an operator, $T^{*}$ is the target of the proposed edit, and $p\in[0,1]$ is the diagnoser confidence.
The target format depends on the operator:
\[
T^{*}=
\begin{cases}
v, & op=\textsc{Split},\\
S, & op=\textsc{Merge},\\
(u,v), & op=\textsc{Update},
\end{cases}
\]
where $v$ is a candidate conflated unit, $S$ is a redundant visible set, and $(u,v)$ denotes an ordered current--superseded pair.
The diagnosis stage does not modify the memory bank.

\subsection{Confidence Gating and Queue Construction}
\label{app:atc:gating}

Each operator has a confidence threshold $\theta_{op}$.
A diagnosis proposal $(op,T^{*},p)$ is accepted only when its confidence satisfies $p\ge\theta_{op}$.
Accepted targets are then passed to \textsc{Norm}, which canonicalizes ids, removes invalid targets, and deduplicates repeated proposals before inserting them into the operator queue $\mathcal{Q}_{op}$.
In our default configuration, we use a shared conservative threshold of $0.9$ for \textsc{Split}, \textsc{Merge}, and \textsc{Update}.
This setting favors high-precision edits: false-positive topology edits are more harmful than delayed consolidation, since an incorrect merge, split, or update can degrade the visible retrieval surface.

\paragraph{Normalization.}
\textsc{Norm} is deterministic and fail-closed.
For \textsc{Split}, the target must be a single visible unit.
For \textsc{Merge}, the target must contain at least two visible units.
For \textsc{Update}, the current unit must be visible, while the superseded unit may be visible or already archived.
Targets that fail these checks are dropped before editing.

\subsection{Serial Editing and Applicability Checks}
\label{app:atc:applicable}

Although diagnosis is parallel, editing is executed serially in the fixed order \textsc{Split}$\rightarrow$\textsc{Merge}$\rightarrow$\textsc{Update}.
This order first separates conflated evidence, then removes redundancy, and finally refreshes versioned descriptors.
Serial execution also makes conflicts deterministic.

During one consolidation run, the executor maintains a set of already modified units.
Before invoking the planner for a queued target, \textsc{Applicable} rechecks three conditions:
\begin{enumerate}[leftmargin=*, nosep]
    \item all referenced unit ids still exist;
    \item the operator-specific visibility and arity constraints still hold;
    \item no referenced unit has already been modified or archived earlier in the same run.
\end{enumerate}
If any condition fails, the target is skipped as stale.
The executor does not repair or retarget stale proposals.

\subsection{Plan Invocation and Fail-Closed Execution}
\label{app:atc:plans}

For each applicable target, ATC invokes the corresponding planning prompt $P_{\mathrm{plan}}$ and obtains a structured repair plan:
\begin{equation}
R \leftarrow \mathrm{LLM}\!\left(P_{\mathrm{plan}}(op,T^{*};\mathcal{M})\right).
\end{equation}
The plan $R$ contains only high-level editing content, such as evidence segments for \textsc{Split} or regenerated descriptors for \textsc{Merge} and \textsc{Update}.
It does not directly edit the graph.
Graph updates are applied only through the deterministic executor:
\begin{equation}
(\mathcal{V},\mathcal{E})
\leftarrow
\textsc{ApplyPlan}(R;\mathcal{V},\mathcal{E}).
\end{equation}

Validation. The executor validates each plan before application.
For \textsc{Split}, the plan must provide a non-empty list of evidence
segments; if fewer than two non-empty segments are returned, the target is
treated as a no-op. Segment descriptors are generated separately with the
descriptor compression prompt. For \textsc{Merge}, the canonical descriptor
must be non-empty and supported by the source units. For \textsc{Update}, the
refreshed descriptor must be non-empty and attached to a visible current unit.
Invalid JSON, missing fields, empty descriptors, or ungrounded segments cause
the target to be skipped.

Fail-closed behavior. ATC performs no retries during consolidation. Each target
ends in one of three states: \textsc{Executed}, \textsc{Skipped}, or
\textsc{Noop}. A target is \textsc{Executed} only if all checks succeed; it is
\textsc{Skipped} if it becomes stale or invalid before planning; and it becomes
\textsc{Noop} if the planner returns a valid plan that declines the edit, e.g.,
a \textsc{Split} plan with fewer than two segments.

\subsection{Determinism}
\label{app:atc:determinism}

ATC is deterministic conditional on the LLM outputs.
All queue routing, target normalization, conflict handling, plan validation, archiving, link insertion, rewiring, and degree-cap pruning are rule-based.
The only model-dependent steps are diagnosis proposal generation and descriptor-level planning.
In implementation, both steps use deterministic decoding and JSON-only outputs, as specified in Appendix~\ref{app:prompts}.

\section{Prompts}
\label{app:prompts}

This appendix lists the effective prompt templates used by ATC.
All prompts require JSON-only outputs and are decoded with temperature $0$.
For compactness, we show the effective prompt templates and expected fields
rather than the full JSON schemas.

The executor validates the returned fields before applying an edit.
Invalid or empty outputs are treated as no-ops at the corresponding planning
stage. Diagnosis outputs are additionally checked by the executor before they
are converted into executable topology proposals.

\subsection{Descriptor Compression Prompt}
\label{app:prompts:descriptor}

The descriptor compression prompt converts raw evidence into a compact semantic
index used for retrieval. It is used when a new memory node is inserted and when
new sibling nodes are created after a split.

\begin{promptbox}{Descriptor Compression Prompt}
\begin{lstlisting}[style=promptstyle]
System: You are a memory compression engine.
Output valid JSON only.

Task:
Compress the following interaction transcript into a concise semantic index.

Input transcript format:
Each line is a message formatted as:
[<message_timestamp>] <role>: <content>

Transcript:
"""
{content}
"""

Conversation reference time:
{timestamp}

Critical requirement: temporal normalization.
1. For each line, treat the timestamp inside brackets as the primary
   reference time for resolving relative temporal expressions.
2. If a line has no valid bracket timestamp, use the conversation
   reference time as fallback.
3. If the content explicitly contains a full absolute date, keep that
   date as-is.
4. Determine one salient absolute date for the whole transcript:
   choose the date that the transcript is primarily about. If multiple
   dates are equally important, choose the most recent date. If no date
   can be resolved, use the date part of the conversation reference time.

Tasks:
1. Write a high-density summary in one or two sentences.
   The summary must start with the salient absolute date in the format:
   [YYYY-MM-DD]
2. Extract three to five critical entities, locations, terms, or topics.

Return JSON:
{
  "summary": "[YYYY-MM-DD] ...",
  "keywords": ["...", "..."]
}
\end{lstlisting}
\end{promptbox}

\paragraph{Expected fields.}
The output contains a \texttt{summary} and a list of \texttt{keywords}.
If either field is missing, the executor falls back to conservative defaults
derived from the raw evidence.

\subsection{Diagnosis Prompt}
\label{app:prompts:diagnosis}

The diagnosis prompt proposes candidate topology edits from a local retrieved
context. It does not modify the memory bank.

\begin{promptbox}{Diagnosis Prompt}
\begin{lstlisting}[style=promptstyle]
Role: Senior Memory Archivist.

Task:
Maintain the integrity of the knowledge graph by identifying redundancy
(Merge), mixed-topic noise (Split), or superseded facts (Update).

Goal:
Group related information and separate distinct topics to improve retrieval
density while avoiding destructive edits.

Data:
[Retrieved Context Nodes: Existing Memory]
{context_text}

[Target Input]
{target_info}

Strict diagnosis rules:

1. Split analysis.
   Trigger:
   Split only if a node contains multi-domain noise that ruins retrieval
   precision.

   Example: split
   "I learned PyTorch today. Also, I need to buy eggs."
   Domains: coding vs. shopping.

   Example: keep
   "I went to the library, studied PyTorch, and then wrote some code."
   This is a coherent narrative flow.

   Constraints:
   - Do not split chronological sequences, user stories, or cause-and-effect
     chains.
   - Do not split a single turn of dialogue if doing so breaks the context of
     who said what.
   - If the content is coherent as a single paragraph, confidence must be 0.0.

2. Merge analysis.
   Trigger:
   Merge only if nodes are factually redundant.

   Example: merge
   Node A: "I live in Tokyo."
   Node B: "My residence is in Tokyo."

   Example: keep
   Node A: "I felt sad yesterday."
   Node B: "I feel happy today."

   Constraints:
   - Check timestamps. Do not merge conflicting states from different times.
   - Ensure the nodes refer to the exact same entity or event.
   - Each merge task may include at most four node ids. If more than four nodes
     are redundant, output multiple merge tasks, each with two to four ids,
     grouped by strongest redundancy.

3. Update analysis.
   Trigger:
   Update only if the target node explicitly contradicts or supersedes an
   existing node.

   Example: update
   Old: "I like apples."
   Target: "I hate apples now."

   Example: update
   Old: "I live in New York."
   Target: "I moved to San Francisco."

   Example: keep
   Old: "I like apples."
   Target: "I like oranges."

   Constraints:
   - The conflict must be explicit.
   - Do not update if the target is merely topically similar to the old node.
   - The old node will be archived and linked as a previous version of the
     target.

Return JSON:
{
  "split_tasks": [
    {
      "node_id": "ID_...",
      "reason": "Specific domains identified: ...",
      "confidence": 0.95
    }
  ],
  "merge_tasks": [
    {
      "node_ids": ["ID_1", "ID_2"],
      "reason": "Redundant static fact regarding ...",
      "confidence": 0.95
    }
  ],
  "update_tasks": [
    {
      "old_node_id": "ID_of_outdated_node",
      "new_node_id": "TARGET",
      "reason": "The target explicitly supersedes the old node.",
      "confidence": 0.95
    }
  ]
}
\end{lstlisting}
\end{promptbox}

\paragraph{Expected fields.}
The output contains three task lists: \texttt{split\_tasks},
\texttt{merge\_tasks}, and \texttt{update\_tasks}. Each task contains target
ids, a short reason, and a confidence score in $[0,1]$.
Only tasks that pass id checks and the high-confidence threshold are converted
into proposal triples $(op,T^{*},p)$ used by ATC.

\subsection{\textsc{Split} Planning Prompt}
\label{app:prompts:split}

The \textsc{Split} planner decomposes one conflated memory unit into grounded
evidence segments. The planner only proposes evidence segments; the executor
creates sibling nodes, derives their descriptors, and rewires links.

\begin{promptbox}{\textsc{Split} Planning Prompt}
\begin{lstlisting}[style=promptstyle]
System: You are an exacting memory editor.
Output valid JSON only.

Task:
Split the target memory into the smallest number of self-contained memory
units, but only when clearly necessary.

Target evidence:
{target_evidence}

Reason for split:
{reason}

Rules:
1. Split only when the evidence contains clearly separable facts, topics, or
   events.
2. Each segment must be copied from the target evidence. Do not invent or
   paraphrase evidence.
3. Preserve speaker tags and temporal cues when present.
4. If the target is coherent as one unit, return exactly one segment identical
   to the original target evidence.
5. Use the smallest number of segments that resolves conflation.

Return JSON:
{
  "segments": ["...", "..."]
}
\end{lstlisting}
\end{promptbox}

\paragraph{Validation.}
The executor requires \texttt{segments} to be a non-empty list of strings.
If fewer than two non-empty segments are returned, the split is treated as a
no-op. Each segment must be grounded in the source evidence. For each accepted
segment, the executor creates a sibling node, derives its summary and keywords
with the descriptor compression prompt, rewires bounded temporal and semantic
edges, adds lineage edges, and archives the parent node.

\subsection{\textsc{Merge} Planning Prompt}
\label{app:prompts:merge}

The \textsc{Merge} planner produces a canonical descriptor for a redundant
visible set. Raw source evidence is not deleted; source nodes are archived and
kept recoverable through lineage links.

\begin{promptbox}{\textsc{Merge} Planning Prompt}
\begin{lstlisting}[style=promptstyle]
System: You are a conservative memory curator.
Output valid JSON only.

Task:
Create one canonical descriptor for a set of redundant memory nodes.

Source memories:
{source_memories}

Reason for merge:
{reason}

Rules:
1. Use only information supported by the source memories.
2. Consolidate duplicated or overlapping facts.
3. Do not merge conflicting states or time-varying preferences.
4. Keep the summary compact but specific.
5. Keywords must be grounded in the source memories.

Return JSON:
{
  "summary": "...",
  "keywords": ["...", "..."]
}
\end{lstlisting}
\end{promptbox}

\paragraph{Validation.}
The executor invokes this prompt only after the candidate set passes validity,
visibility, and semantic-coherence checks. It requires a non-empty
\texttt{summary} and a non-empty list of \texttt{keywords}. Keywords are
deterministically normalized by removing duplicates and empty strings. If the
descriptor is invalid or empty after normalization, the merge is treated as a
no-op.

\subsection{\textsc{Update} Planning Prompt}
\label{app:prompts:update}

The \textsc{Update} planner refreshes the descriptor of a current memory unit
while marking an older unit as superseded. The superseded unit can be used to
identify stale or redundant details, but it is not a source of new facts for the
current descriptor.

\begin{promptbox}{\textsc{Update} Planning Prompt}
\begin{lstlisting}[style=promptstyle]
System: You are an exacting memory updater.
Output valid JSON only.

Task:
Refresh the descriptor of the current memory node using the current evidence
and the superseded memory.

Current node evidence:
{current_evidence}

Current summary:
{current_summary}

Current keywords:
{current_keywords}

Superseded node evidence:
{superseded_evidence}

Superseded summary:
{superseded_summary}

Superseded keywords:
{superseded_keywords}

Reason for update:
{reason}

Rules:
1. Updated fields must be supported by the current evidence.
2. Use the superseded node only to identify outdated or redundant details.
3. If current and superseded evidence conflict, prioritize the current
   evidence.
4. Improve retrieval clarity by disambiguating entities, time, and topic.
5. Do not introduce unsupported facts.

Return JSON:
{
  "updated_summary": "...",
  "updated_keywords": ["...", "..."]
}
\end{lstlisting}
\end{promptbox}

\paragraph{Validation.}
The executor requires a non-empty \texttt{updated\_summary} and a non-empty
list of \texttt{updated\_keywords} before rewriting the current descriptor.
If the refreshed descriptor is invalid or empty, the descriptor rewrite is
treated as a no-op. When the descriptor is unchanged after normalization, only
the descriptor rewrite is skipped; the accepted supersession edit may still
archive the older node and add the version link.

\begin{table}[ht]
\centering
\footnotesize
\setlength{\tabcolsep}{4pt}
\caption{Dataset statistics for the evaluated benchmarks.}
\label{tab:dataset_stats}
\begin{tabular}{lcc}
\toprule
\textbf{Statistic} & \textbf{LoCoMo} & \textbf{LongMemEval-s} \\
\midrule
Conversations & 10 & 500 \\
Avg. sessions / conv. & 27.2 & 50.2 \\
Avg. queries / conv. & 198.6 & 1.0 \\
Avg. tokens / conv. & 20.1k & 103.1k \\
Session dates & Yes & Yes \\
Retrieval labels & Yes & Yes \\
QA labels & Yes & Yes \\
Dialogue type & User--User & User--AI \\
\bottomrule
\end{tabular}
\end{table}

\section{Experimental Details and Fairness}
\label{app:exp_details}

This appendix provides the implementation and evaluation details used in our experiments.
All methods are evaluated in chronological interaction order and answer each query once.
We do not train or fine-tune any model parameters.

\subsection{Benchmarks, Statistics, and Metrics}
\label{app:exp:benchmarks}

We evaluate on LoCoMo and LongMemEval-s.
Table~\ref{tab:dataset_stats} summarizes their basic statistics.
For answer quality, we report GPT-4o-as-judge score (4o-J), F1, BLEU-1, and ROUGE-L.
For retrieval quality, we report Recall@5 (R@5) and NDCG@5 (N@5), computed using benchmark-provided evidence annotations.

The two benchmarks use different retrieval granularities.
LoCoMo uses turn-level evidence matching, while LongMemEval-s uses session-level evidence matching.
Therefore, retrieval scores are not directly comparable across datasets; we interpret them together with downstream answer quality.

\subsection{Models and Shared Configuration}
\label{app:exp:models}

All methods use the same answer generator, GPT-4o-mini, with temperature $0$.
Unless otherwise stated, all All-Mem LLM calls, including online descriptor generation, ATC diagnosis, and operator planning, also use GPT-4o-mini with temperature $0$.
All-Mem uses All-MiniLM-L6-v2 to embed memory descriptors and queries.
Cosine similarity is used for dense retrieval and re-ranking.

For All-Mem, Stage~1 retrieves anchors from the visible surface $\mathcal{V}^{+}$.
Stage~2 expands over typed links under a fixed hop and candidate budget.
Stage~3 ranks candidates by query--memory embedding similarity and injects the selected evidence into the generator context.

\begin{table}[t]
\centering
\footnotesize
\setlength{\tabcolsep}{4pt}
\caption{Default configuration for All-Mem in the main experiments.}
\label{tab:default_config}
\begin{tabular}{ll}
\toprule
\textbf{Component} & \textbf{Setting} \\
\midrule
Generator & GPT-4o-mini, temperature $0$ \\
All-Mem LLM calls & GPT-4o-mini, temperature $0$ \\
Embedding model & All-MiniLM-L6-v2 \\
Stage-1 anchors & $k=10$ \\
Hop budget & $H_q=4$ \\
Expansion budget & $L=40$ \\
Final injected units & $K=16$ \\
Semantic out-degree cap & $d_{\sigma}=8$ \\
ATC threshold & $\theta=0.9$ \\
Operator order & \textsc{Split}$\rightarrow$\textsc{Merge}$\rightarrow$\textsc{Update} \\
Offline trigger & Every 3 sessions \\
\bottomrule
\end{tabular}
\end{table}

\subsection{Compared Methods}
\label{app:exp:baselines}

We compare All-Mem with Full History, Naive RAG, MemGPT, A-Mem, HippoRAG2, Mem0, and LightMem.
All baselines are run under the same benchmark order and the same answer generator whenever applicable.
For memory-centric baselines, we preserve their intended memory construction and retrieval procedures, including their native memory-writing or maintenance steps.

Naive RAG uses dense retrieval over accumulated memory units.
Full History injects the available conversation history under the same generator-side context constraint.
For methods that return ranked evidence, we preserve their native ranking and apply the same final context-budget constraint before answer generation.

\begin{table*}[ht]
\centering
\footnotesize
\setlength{\tabcolsep}{4pt}
\caption{
\textbf{Full ablation results on LoCoMo and LongMemEval-s.}
Answer quality is measured by 4o-J and F1; retrieval quality is measured by R@5 and N@5.
For LongMemEval-s, retrieval is session-level and should be interpreted together with downstream QA.
}
\label{tab:ablation_full}
\begin{tabular}{lcccccccc}
\toprule
\textbf{Variant}
& \multicolumn{4}{c}{\textbf{LoCoMo}}
& \multicolumn{4}{c}{\textbf{LongMemEval-s}} \\
\cmidrule(lr){2-5}\cmidrule(lr){6-9}
& \textbf{4o-J} & \textbf{F1} & \textbf{R@5} & \textbf{N@5}
& \textbf{4o-J} & \textbf{F1} & \textbf{R@5} & \textbf{N@5} \\
\midrule
\textbf{All-Mem (Full)}
& \textbf{54.63} & \textbf{52.18} & \textbf{46.63} & \textbf{41.02}
& \textbf{60.20} & \textbf{45.19} & \textbf{94.68} & \textbf{93.27} \\
\midrule
No-Visibility
& 50.87 & 47.63 & 41.84 & 34.76
& 56.20 & 40.60 & 91.24 & 89.43 \\
\midrule
w/o \textsc{Split}
& 53.21 & 50.94 & 45.12 & 39.58
& 59.00 & 43.30 & 93.76 & 92.15 \\
w/o \textsc{Merge}
& 52.08 & 49.02 & 43.96 & 38.14
& 57.80 & 42.10 & 92.46 & 90.73 \\
w/o \textsc{Update}
& 52.67 & 49.74 & 44.41 & 39.02
& 58.20 & 42.60 & 92.98 & 91.47 \\
\midrule
Anchors-only
& 51.76 & 49.27 & 43.19 & 36.54
& 57.40 & 41.70 & 92.13 & 90.18 \\
No-recovery-links
& 52.74 & 50.06 & 44.07 & 38.43
& 58.40 & 42.80 & 93.04 & 91.24 \\
No-type-priority
& 53.62 & 50.91 & 45.18 & 39.66
& 59.20 & 43.70 & 93.67 & 92.05 \\
\bottomrule
\end{tabular}
\end{table*}

\subsection{Context Budget and Token Accounting}
\label{app:exp:tokens}

To avoid giving any method an advantage from larger generator inputs, all methods are evaluated under the same final generator context budget.
If a retrieval method returns more evidence than fits the budget, we truncate by its native ranking order.
Truncation is applied at the evidence-unit boundary whenever possible.

We report two types of token costs.
Construction tokens count LLM input and output tokens used for memory writing or maintenance.
Retrieval tokens denote the realized generator input tokens contributed by retrieved evidence at query time.
For offline ATC, event-level token costs are amortized over the realized number of turns between offline events.

\subsection{Latency Measurement}
\label{app:exp:latency}

We separately measure online writing latency, offline ATC latency, and query-time memory-module latency.
Online writing latency is measured per interaction turn.
Offline ATC latency is measured per consolidation event and is also reported after amortization over turns.
Query-time memory-module latency includes query embedding, Stage~1 dense retrieval, Stage~2 typed-link expansion, and Stage~3 re-ranking.
It excludes generator inference and decoding.

All latency measurements are collected with cached memory embeddings and a warm retrieval pipeline.
Index rebuilding or offline consolidation is reported separately and is not included in query-time memory-module latency.

\subsection{Fairness Controls}
\label{app:exp:fairness}

We use the following controls to ensure that performance differences reflect memory organization rather than evaluation artifacts:
\begin{itemize}[leftmargin=*, nosep]
    \item all methods process sessions chronologically;
    \item all methods use the same answer generator and decoding temperature;
    \item all methods answer each query once;
    \item retrieval and QA metrics are computed with the same evaluation scripts;
    \item final generator context size is matched across methods;
    \item token and latency costs are logged separately for memory construction, offline maintenance, retrieval, and generation.
\end{itemize}
These controls preserve baseline-specific memory mechanisms while keeping answer generation and evaluation comparable.

\section{Full Ablation Study}
\label{app:ablation_full}

This appendix provides the full ablation results for All-Mem.
The main text reports a representative subset of variants; here we include all variants evaluated on both LoCoMo and LongMemEval-s.
Unless otherwise stated, all variants use the same generator, embedding model, ATC threshold, retrieval budgets, and evaluation protocol as the full model.
Each variant removes or modifies one component while keeping the rest of the pipeline unchanged.

\subsection{Ablation Variants}
\label{app:ablation:variants}

\paragraph{No-Visibility.}
This variant removes visible-surface gating by forcing all units to remain searchable.
Thus, Stage~1 anchoring searches the entire memory bank rather than the maintained visible surface.
The drop in both QA and retrieval quality shows that exposing all accumulated units increases competition from redundant, stale, or noisy evidence under a fixed retrieval budget.

\paragraph{Operator ablations.}
The variants w/o \textsc{Split}, w/o \textsc{Merge}, and w/o \textsc{Update} disable one topology operator at a time during offline ATC.
Removing \textsc{Split} weakens fine-grained localization because conflated evidence remains entangled.
Removing \textsc{Merge} keeps redundant variants visible, lowering the information density of the searchable surface.
Removing \textsc{Update} weakens handling of superseded states, making stale evidence more likely to remain competitive during retrieval.

\paragraph{Retrieval-structure ablations.}
\textbf{Anchors-only} disables Stage~2 typed-link expansion and ranks only the visible anchors returned by Stage~1.
\textbf{No-recovery-links} keeps the memory topology but prevents traversal over recovery-oriented links during expansion.
\textbf{No-type-priority} removes the type-aware traversal priority and expands eligible links in a uniform deterministic order.
The performance drops show that All-Mem benefits not only from a curated visible surface, but also from controlled expansion to linked evidence.

\subsection{Discussion}
\label{app:ablation:discussion}

The ablations support three conclusions.
First, visible-surface gating is necessary: searching the full bank degrades selectivity under fixed budgets.
Second, all three topology operators contribute to performance, with \textsc{Merge} having the largest effect because redundancy is frequent in long-horizon interaction histories.
Third, typed-link expansion improves evidence coverage beyond nearest visible anchors, while type priority provides a smaller but consistent gain in allocating the expansion budget.
Overall, the full system performs best because it combines compact visible-surface maintenance with bounded recovery to archived evidence.

\section{Efficiency and Scalability}
\label{app:efficiency}

This appendix provides additional diagnostics for the efficiency claims in Sec.~\ref{subsec:cost_analysis}.
The main text reports the overall online, offline, and query-time cost profile.
Here we focus on why query-time retrieval remains scalable: Stage~1 searches the maintained visible surface $\mathcal{V}_N^{+}$ instead of the full memory bank $\mathcal{V}_N$, while Stage~2 and Stage~3 are explicitly bounded by the hop and candidate budgets.

\subsection{Measurement Protocol}
\label{app:efficiency:protocol}

We measure memory-module latency (\emph{Mem-Lat}) as the end-to-end runtime of query-time retrieval, excluding generator inference and decoding.
It includes query embedding, Stage~1 dense retrieval, Stage~2 hop-bounded typed-link expansion, and Stage~3 re-ranking.
All measurements use the same embedding model and fixed retrieval budgets as the main experiments.

Stage~1 computes cosine similarities between a query embedding and a cached embedding matrix, then extracts the Top-$k$ candidates by partial sorting.
Stage~2 expands over typed links under hop budget $H_q$ and candidate budget $L$.
Stage~3 re-ranks at most $L$ candidates before selecting the final evidence units for generator input.

\subsection{Visible-Surface Dynamics}
\label{app:efficiency:visible}

A core design goal of All-Mem is to keep the searchable surface compact as histories grow.
Fig.~\ref{fig:efficiency_pack_app}(a) tracks the visible surface size $|\mathcal{V}_N^{+}|$ across consolidation checkpoints.
An append-only memory that keeps all units searchable follows the reference line $|\mathcal{V}_N^{+}|=N$.
In contrast, All-Mem periodically archives redundant or superseded units through non-destructive consolidation, so Stage~1 anchoring is governed by the smaller maintained surface rather than by the full accumulated bank.

\subsection{Stage-Wise Latency}
\label{app:efficiency:stagewise}

Fig.~\ref{fig:efficiency_pack_app}(b) decomposes Mem-Lat into the three retrieval stages.
Stage~2 expansion is lightweight because it is both hop-bounded and candidate-bounded.
The dominant components are Stage~1 dense retrieval and Stage~3 re-ranking, which depend on the searchable surface and candidate budget, respectively.
This supports the design choice of using offline consolidation to control the visible surface before query-time retrieval.

\subsection{Search-Space Scaling}
\label{app:efficiency:scaling}

To isolate the effect of visible-surface search, we measure Stage~1 Top-$k$ retrieval latency at consolidation checkpoints while increasing history length $N$.
For each checkpoint, we compare dense cosine retrieval over the full bank $\mathcal{V}_N$ with retrieval over the visible surface $\mathcal{V}_N^{+}$, using the same implementation and the same $k$.

\begin{figure*}[t]
  \centering
  \begin{minipage}{0.32\textwidth}
    \centering
    \includegraphics[width=\linewidth,trim=8 6 8 6,clip]{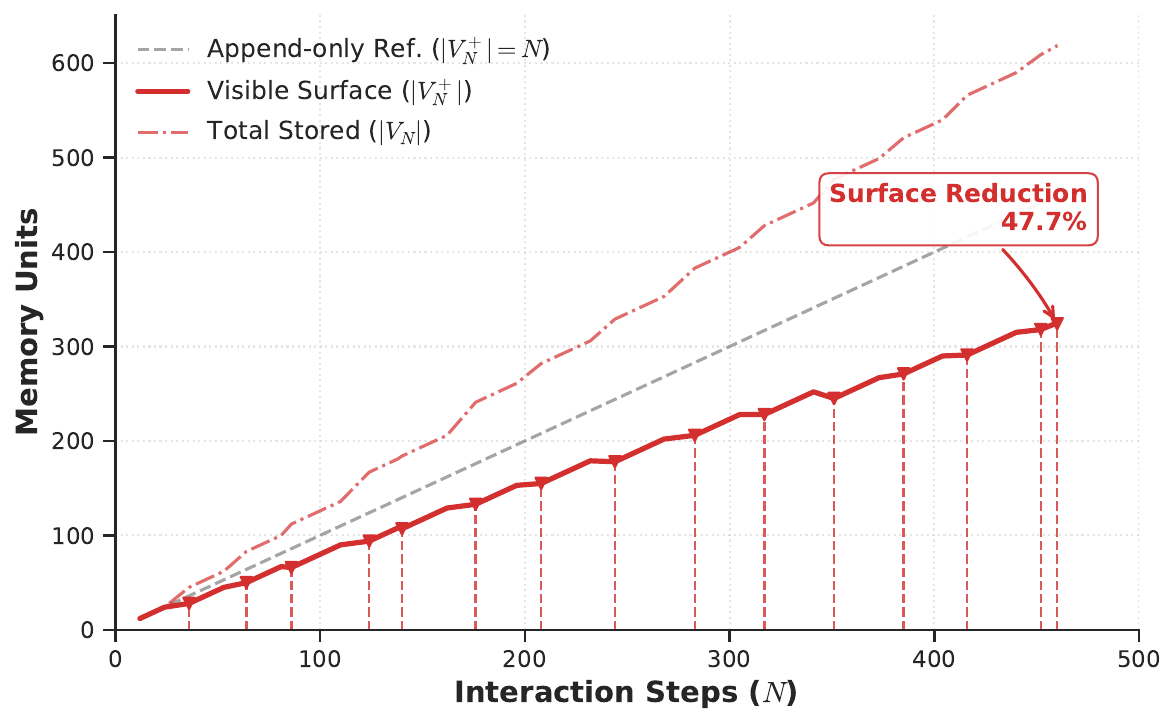}
    \vspace{-1mm}
    \caption*{\small (a) Visible-surface dynamics.}
  \end{minipage}
  \hfill
  \begin{minipage}{0.32\textwidth}
    \centering
    \includegraphics[width=\linewidth,trim=8 6 8 6,clip]{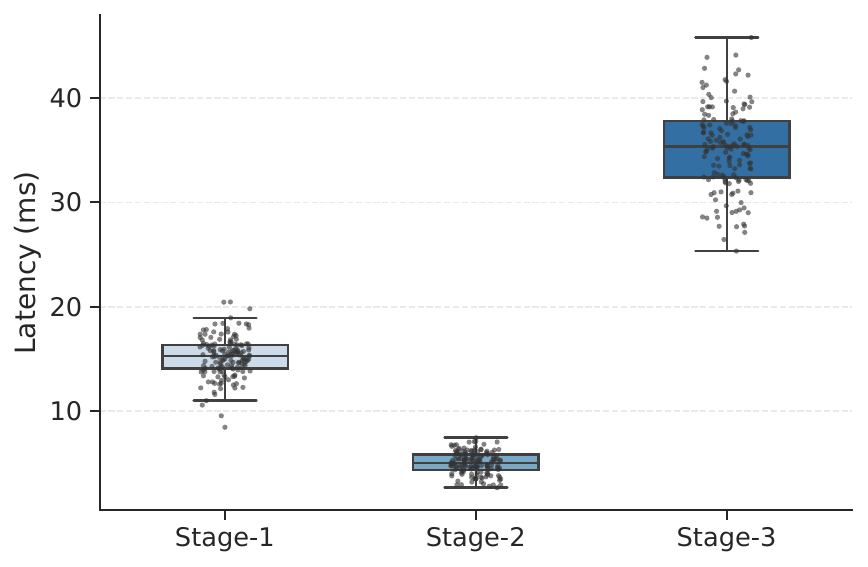}
    \vspace{-1mm}
    \caption*{\small (b) Stage-wise latency.}
  \end{minipage}
  \hfill
  \begin{minipage}{0.32\textwidth}
    \centering
    \includegraphics[width=\linewidth,trim=8 6 8 6,clip]{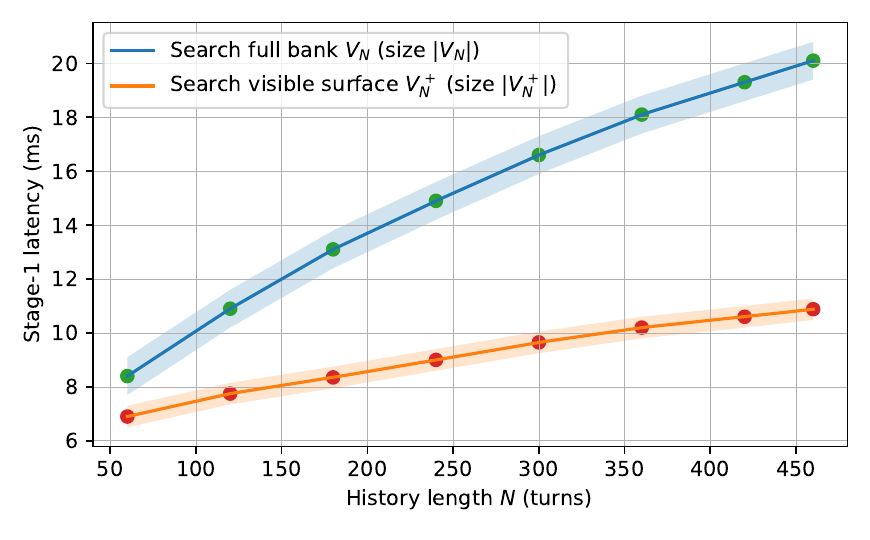}
    \vspace{-1mm}
    \caption*{\small (c) Stage-1 scaling.}
  \end{minipage}
  \vspace{-1mm}
  \caption{\textbf{Efficiency diagnostics.}
  (a) The searchable visible surface $|\mathcal{V}_N^{+}|$ grows more slowly than the append-only reference $|\mathcal{V}_N^{+}|=N$.
  (b) Query-time memory latency is decomposed into Stage~1 dense retrieval, Stage~2 typed-link expansion, and Stage~3 re-ranking.
  (c) Stage~1 Top-$k$ latency when searching over the full bank $\mathcal{V}_N$ versus the visible surface $\mathcal{V}_N^{+}$ at consolidation checkpoints; shaded bands denote interquartile ranges.}
  \label{fig:efficiency_pack_app}
  \vspace{-5pt}
\end{figure*}

As shown in Fig.~\ref{fig:efficiency_pack_app}(c), searching over $\mathcal{V}_N^{+}$ consistently reduces Stage~1 latency as history length grows.
At $N=460$, visible-surface search reduces Stage~1 time from 20.10ms to 10.88ms, a 45.9\% reduction.
This confirms that All-Mem's query-time retrieval cost is governed by the maintained visible surface rather than the unbounded history length.

\subsection{Implications}
\label{app:efficiency:implications}

The diagnostics support three conclusions.
First, offline consolidation keeps the searchable surface smaller than the append-only history.
Second, Stage~2 expansion adds little query-time overhead because traversal is explicitly bounded.
Third, visible-surface anchoring reduces Stage~1 latency without deleting archived evidence, since archived units remain recoverable through typed links.
Together with the cost profile in Sec.~\ref{subsec:cost_analysis}, these results support the intended online/offline separation of All-Mem.

\section{Recoverability Analysis}
\label{app:recoverability}

All-Mem archives outdated or redundant units instead of deleting them.
This appendix evaluates whether archived units remain reachable from the current visible surface through the maintained typed-link topology.

\subsection{Metric}
\label{app:recoverability:metric}

We evaluate bank-induced recoverability on the final consolidated snapshot of LongMemEval-s.
For each archived unit $v\in\mathcal{V}_N\setminus\mathcal{V}_N^{+}$, we compute its directed hop distance from the visible surface:
\begin{equation}
h(v)=
\min_{u\in\mathcal{V}_N^{+}}
\mathrm{dist}_{\mathcal{E}_N}(u,v),
\end{equation}
where $\mathrm{dist}_{\mathcal{E}_N}$ denotes the directed shortest-path distance induced by the typed-link graph $\mathcal{E}_N$.
If no directed path exists, we set $h(v)=\infty$.

Given a hop budget $H$, we define budgeted recoverability as the fraction of archived units reachable within $H$ hops:
\begin{equation}
\mathrm{Cov}(H)=
\frac{
|\{v\in\mathcal{V}_N\setminus\mathcal{V}_N^{+}:h(v)\le H\}|
}{
|\mathcal{V}_N\setminus\mathcal{V}_N^{+}|
}.
\end{equation}
This metric measures whether non-visible evidence remains accessible to hop-bounded expansion after consolidation.

\begin{figure}[t]
    \centering
    \includegraphics[width=0.95\linewidth]{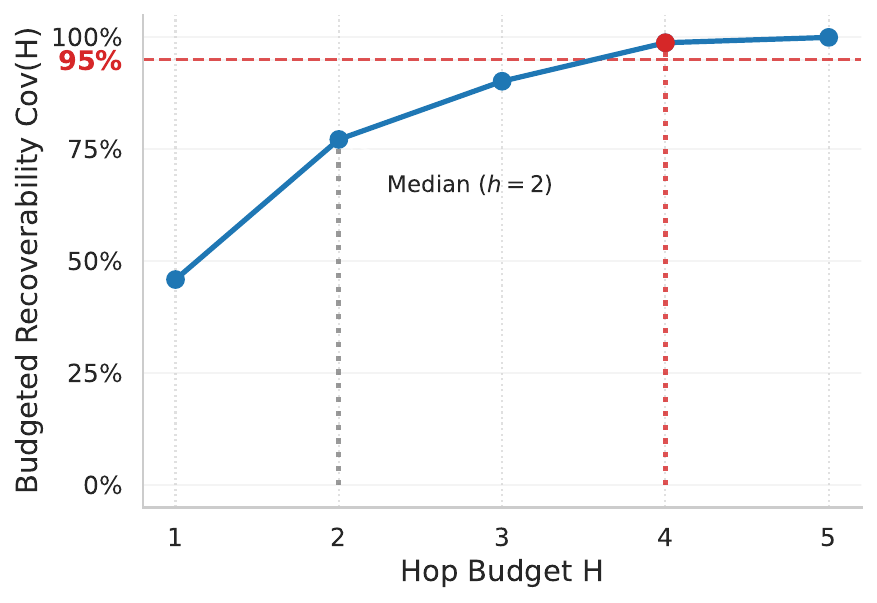}
    \caption{\textbf{Budgeted recoverability analysis.}
    Coverage $\mathrm{Cov}(H)$ of archived units reachable within hop budget $H$ over the directed typed-link topology $\mathcal{E}_N$.
    The dashed line marks the hop budget achieving 95\% coverage ($H_{0.95}=4$); the median distance is $H=2$.}
    \label{fig:recoverability}
    \vspace{-4pt}
\end{figure}

\subsection{Results}
\label{app:recoverability:results}

Fig.~\ref{fig:recoverability} shows that archived evidence remains shallowly reachable.
On the final LongMemEval-s snapshot, all archived units are reachable from the visible surface, yielding 0\% unreachable archived units.
The median archived-unit distance is $2$ hops, and 95\% coverage is achieved at $H_{0.95}=4$.

These results support the intended role of non-destructive consolidation.
Archiving removes noisy, redundant, or superseded units from the searchable surface, but version and sibling links preserve short directed paths back to the original evidence.
Thus, query-time retrieval can keep Stage~1 selective while still recovering archived evidence through bounded Stage~2 expansion when needed.

\section{Edit Quality Audit}
\label{app:edit_quality}

All-Mem relies on LLM-generated diagnosis and planning outputs to propose topology edits.
To directly assess whether these edits are semantically appropriate, we manually audit a stratified sample of accepted ATC edits on LongMemEval-s.
The audit focuses on edit correctness rather than downstream answer quality.

\subsection{Annotation Protocol}
\label{app:edit_quality:protocol}

We sample accepted \textsc{Split}, \textsc{Merge}, and \textsc{Update} edits after confidence gating and executor validation.
Each edit is labeled as \emph{valid} if it correctly identifies the intended structural drift pattern and applies an appropriate non-destructive change:
\textsc{Split} should separate genuinely conflated evidence,
\textsc{Merge} should consolidate factually redundant units,
and \textsc{Update} should archive a genuinely superseded state while preserving the current one.

We additionally label an edit as \emph{harmful} if it is likely to degrade future retrieval.
Typical harmful cases include over-splitting a coherent memory, merging units that share surface entities but contain distinct facts, or treating complementary information as a superseding update.
Because All-Mem is non-destructive, harmful edits do not delete raw evidence; however, they can still affect which units remain on the visible surface.

\begin{table}[t]
\centering
\footnotesize
\setlength{\tabcolsep}{5pt}
\caption{
\textbf{Human audit of accepted ATC edits.}
``Audit'' is the number of accepted edits manually inspected.
``Valid'' is the fraction judged semantically correct.
``Harmful'' is the fraction judged likely to degrade future retrieval.
}
\label{tab:edit_quality}
\begin{tabular}{lccc}
\toprule
\textbf{Op.} 
& \textbf{Audit} 
& \textbf{Valid} 
& \textbf{Harmful} \\
\midrule
\textsc{Split}  & 40 & 90.0\% & 5.0\% \\
\textsc{Merge}  & 80 & 93.8\% & 3.8\% \\
\textsc{Update} & 70 & 95.7\% & 2.9\% \\
\midrule
\textbf{All}    & 190 & 94.2\% & 3.7\% \\
\bottomrule
\end{tabular}
\end{table}

\subsection{Results}
\label{app:edit_quality:results}

Table~\ref{tab:edit_quality} shows that accepted ATC edits have high semantic precision after confidence gating and executor-side validation.
\textsc{Update} achieves the highest valid rate, since supersession is often indicated by explicit contradictions, temporal cues, or user corrections.
\textsc{Merge} is also reliable, but its main risk is merging memories that appear similar because they mention the same entity while actually encoding different facts.
\textsc{Split} is the most difficult operation, because the boundary between a genuinely conflated memory and a broad but coherent memory can be ambiguous.

The harmful-edit rate remains low across operators.
The most common harmful patterns are:
(i) over-splitting coherent multi-sentence memories,
(ii) merging semantically related but non-redundant facts,
and (iii) treating complementary details as temporal supersession.
These errors explain why ATC uses a conservative confidence threshold and fail-closed executor validation.

\begin{figure*}[ht]
  \centering
  \includegraphics[width=0.92\textwidth]{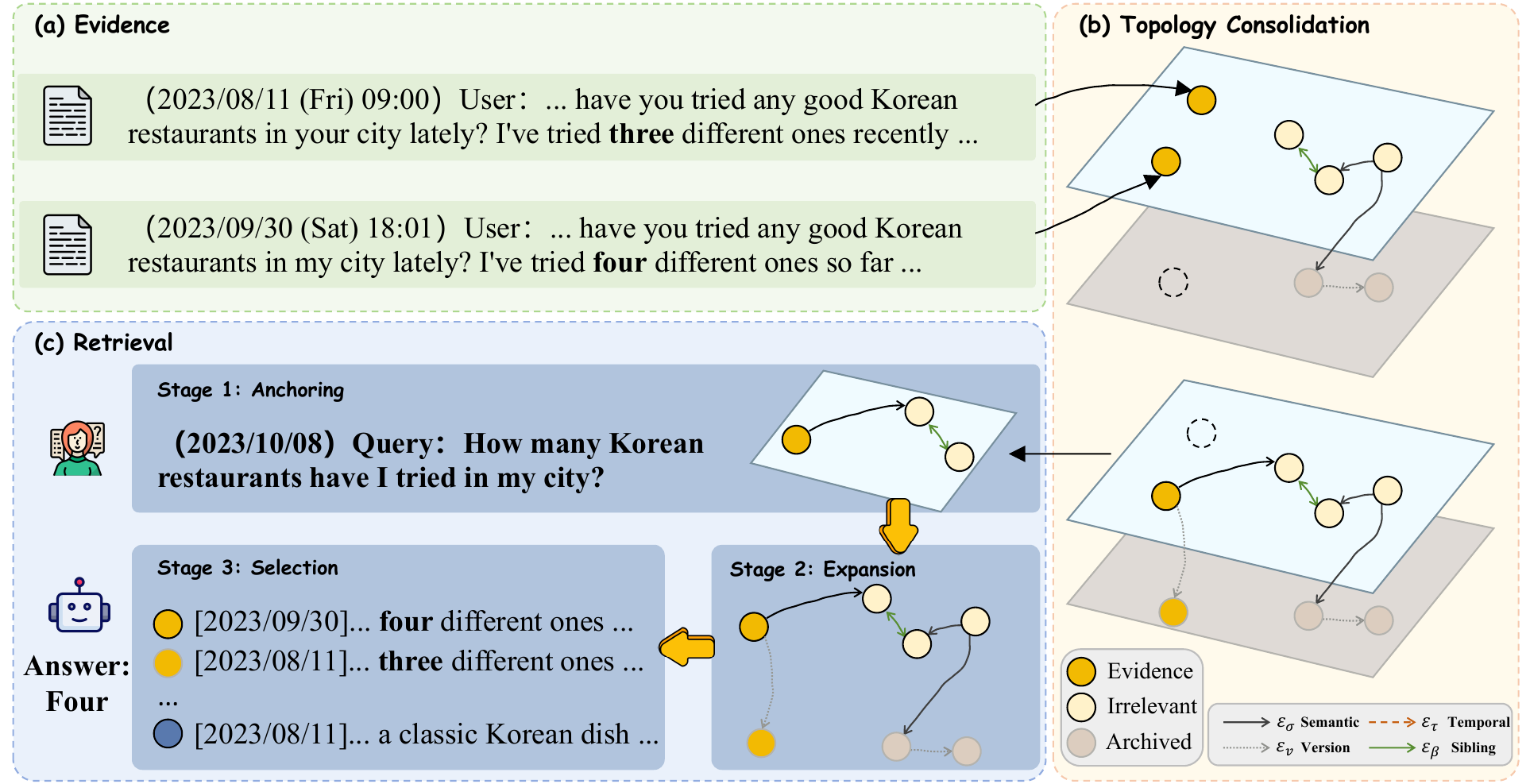}
  \vspace{-2mm}
  \caption{\textbf{Case study: \textsc{Update} with budgeted recovery.}
  A user first states that they tried \emph{three} Korean restaurants, and later updates the fact to \emph{four}.
  (a) Before consolidation, both the older and newer evidence units may compete during retrieval.
  (b) ATC applies \textsc{Update}: the newer unit remains on the visible surface $\mathcal{V}_N^{+}$, while the superseded unit is archived and connected through a version link.
  (c) At query time, Stage~1 anchors on the visible surface and retrieves the current fact; Stage~2 can still recover the archived predecessor through typed-link expansion if supporting history is needed.
  Stage~3 prioritizes the up-to-date evidence, yielding the correct answer under the context budget.}
  \label{fig:case_update}
  \vspace{-6pt}
\end{figure*}

\section{Case Study}
\label{app:case_study}

This appendix provides a qualitative example illustrating how All-Mem handles superseded evidence under a fixed retrieval budget.
The case focuses on \textsc{Update}, since temporal supersession is a common failure mode in long-horizon memory: an older fact can remain highly similar to the query even after the user has provided a newer correction.

\paragraph{Superseded-state failure.}
The example in Fig.~\ref{fig:case_update} shows a user preference or experience that changes over time.
Before consolidation, both the old and new statements can be semantically close to the same query.
A flat retriever may therefore surface the obsolete statement together with, or instead of, the current one, causing temporally inconsistent answers.

\paragraph{Topology consolidation.}
All-Mem resolves this case by applying \textsc{Update}.
The newer unit is kept visible and its descriptor is refreshed for retrieval, while the superseded unit is archived by setting its visibility indicator to zero.
A directed version link from the current unit to the archived unit preserves provenance without allowing the older statement to remain a first-class search target on the visible surface.

\paragraph{Query-time behavior.}
At query time, Stage~1 searches only $\mathcal{V}_N^{+}$, so it preferentially anchors on the current fact.
If the query requires historical support, Stage~2 can recover the archived predecessor by following typed links within the hop budget.
Thus, the system is both \emph{selective}, because stale evidence no longer crowds the initial retrieval surface, and \emph{recoverable}, because the original evidence remains accessible.

\end{document}